\documentclass{amsart}

\usepackage{latexsym}
\usepackage{amsopn}
\usepackage{pstricks}
\usepackage{graphicx}
\usepackage{amsfonts,amsmath,amsthm,amssymb,mathrsfs,verbatim}

\numberwithin{equation}{section}

\newtheorem{lemma}{Lemma}[section]
\newtheorem{theorem}{Theorem}[section]

\newtheorem{proposition}{Proposition}[section]

\newcommand{\qbin}[2]{\genfrac{[}{]}{0pt}{}{#1}{#2}}
\newcommand{\abs}[1]{\lvert#1\rvert}

\newcommand{\B}{\mathcal B}
\newcommand{\Bup}{\textup{B}}
\newcommand{\Bupp}{\textup{B'}}
\newcommand{\Fup}{\textup{F}}
\newcommand{\Sup}{\textup{S}}
\newcommand{\Z}{\mathbb Z}
\newcommand{\Vh}{\hat{V}}
\renewcommand{\Mc}{\mathcal M}
\newcommand{\la}{\lambda}
\newcommand{\Ra}{\rangle}

\newcommand{\osph}{\widehat{\textup{osp}}}
\newcommand{\uh}{\widehat{\textup{u}}}
\newcommand{\suh}{\widehat{\textup{su}}}
\newcommand{\ppmod}[1]{~(\!\bmod#1)}

\begin{document}

\title[Graded parafermions]{Characters of graded parafermion conformal field theory}

\author[Fortin, Mathieu, Warnaar]
{Jean-Fran\c cois Fortin, Pierre Mathieu and S.~Ole Warnaar}
\thanks{JFF is supported by FQRNT, PM is supported by NSERC and SOW is 
supported by the ARC Centre of Excellence MASCOS}
\address{Department of Physics and Astronomy, Rutgers, The State University
of New Jersey, Piscataway, NJ 08854-8019}
\email{jffor27@physics.rutgers.edu}
\address{D\'epartement de physique, de g\'enie physique et d'optique,
Universit\'e Laval, Qu\'ebec, Canada, G1K 7P4}
\email{pmathieu@phy.ulaval.ca}
\address{Department of Mathematics and Statistics,
The University of Melbourne, Vic 3010, Australia}
\email{warnaar@ms.unimelb.edu.au}

\begin{abstract}
The graded parafermion conformal field theory at level $k$
is a close cousin of the much-studied $\Z_k$ parafermion model.
Three character formulas for the graded parafermion theory
are presented, one bosonic, one fermionic (both previously known) and
one of spinon type (which is new).
The main result of this paper is a proof of the equivalence of these 
three forms using $q$-series methods combined with the combinatorics
of lattice paths. The pivotal step in our approach
is the observation that the graded 
parafermion theory --- 
which is equivalent to the coset $\osph(1,2)_k/\uh(1)$ ---
can be factored as $(\osph(1,2)_k/\suh(2)_k) \times (\suh(2)_k/\uh(1))$, 
with the two cosets on the right equivalent to the minimal model 
${\mathcal M}(k+2,2k+3)$ and the $\Z_k$ parafermion model,
respectively. This factorisation allows for a
new combinatorial description of the 
graded parafermion characters in terms of the
one-dimensional configuration sums of the $(k+1)$-state
Andrews--Baxter--Forrester model.
\end{abstract}

\maketitle

\section{Introduction}

A conformal field theory typically has many different formulations. 
For instance, it may be formulated (i) as the representation 
theory of some extended conformal algebra, (ii) as a coset 
model, (iii) as a free-field representation, (iv) in terms of a
quasi-particle description. 
Moreover, within these different categories there can be more than one 
description, e.g., the model under study may be described in terms 
of an irreducible representation of more than one conformal algebra, 
or have more than one coset representation.
Different constructions of the same irreducible modules may lead to
formally equivalent characters which structurally appear rather 
distinct.
This poses the problem of establishing their direct equivalence. 
When successful, such a verification, in turn, places the physical 
argument underlying the construction of the characters on a sound basis. 

The aim of this work is to provide direct proofs --- at the level
of character representations --- of different algebraic descriptions of
graded parafermion theories.
Since we rely heavily on the non-graded parafermion theory,
we will first summarize current understanding of this more familiar 
and much better understood conformal field theory.

\subsection{Ordinary parafermion characters}
The $\Z_k$ parafermion conformal theory \cite{ZF} provides a good 
illustration of a conformal field theory with a multitude of 
formulations. The theory is defined by the algebra of parafermionic
fields $\psi_1$ and $\psi_1^{\dagger}$ of dimension $1-1/k$ 
and central charge 
$2(k-1)/(k+2)$.
The highest-weight modules are parametrized by an integer (Dynkin label)
$\ell$ with $0\leq \ell<k$.
We are interested in modules with fixed relative charge $2r$, where
$r$ counts the number of $\psi_1$ modes minus the number of
$\psi_1^{\dagger}$ modes.
Suppressing the level $k$, the corresponding characters will be 
denoted by $\hat{\chi}_{\ell,r}(q)$.
Here the `hat' has been used to distinguish ordinary and graded
parafermion characters.

The $\Z_k$ parafermion model is known to be
equivalent to the coset
\begin{equation}\label{cos}
\frac{\suh(2)_k}{\uh(1)}\;.
\end{equation}
The first derivation of an explicit expression for the parafermionic 
characters by Kac and Petersen relied on \eqref{cos} and
identifies the characters with the coset branching functions. 

Let $\hat{\chi}_{\ell}(x;q)$ denote the character of the 
$\suh(2)_k$ integrable module of Dynkin label $\ell$ 
(with $0\leq \ell \leq k$), 
and let $K_m(x;q)$ denote $\uh(1)$ character of charge $m$, 
associated to the current algebra of a boson compactified
on an integer square-radius $R^2=2k$ \cite[Sec. 14.4.4]{CFT}.
Then the branching functions associated to the coset
\eqref{cos} are given by the decomposition 
\begin{equation}\label{bfct}
\hat{\chi}_{\ell}(x;q)=
\sum_{m=1-k}^k \hat{b}_{\ell, m}(q)\, K_m(x;q).
\end{equation}
The above-mentioned identification of branching functions
and parafermion characters implies that for
$m-\ell$ even
\begin{equation}\label{bchihat}
\hat{b}_{\ell,m}(q)=\hat{\chi}_{\ell,\frac{m-\ell}{2}}(q).
\end{equation}
(It can in fact easily be shown that $\hat{b}_{\ell,m}(q)=0$ 
when $m-\ell$ is odd.)

According to the Weyl--Kac formula, the $\suh(2)_k$ characters
are given by a ratio of differences of theta functions as
\begin{equation}\label{WK}
\hat{\chi}_{\ell}(x;q)=
\frac{\Theta^{(k+2)}_{\ell+1}(x;q)-\Theta^{(k+2)}_{-\ell-1}(x;q)}
{\Theta^{(2)}_{1}(x;q)-\Theta^{(2)}_{-1}(x;q)}.
\end{equation}
Using \eqref{bfct}, \eqref{bchihat} and \eqref{WK},
Kac and Petersen \cite{KP} obtained
the following expression for the parafermion characters:
\begin{align}\label{string}
\hat{\chi}_{\ell,r}^{\Bup}(q)=\frac{1}{\eta^2(q)}
\Biggl\{&\Biggl(\;\sum_{\substack{i\geq 0 \\[0.5mm] j\geq 0}} -
\sum_{\substack{i<0\\[0.5mm] j<0}}\;\Biggr) (-1)^i
q^{\bigl(\ell+1+(i+2j)(k+2)\bigr)^2/(4(k+2))-(m+ik)^2/(4k)}\\
- &\Biggl(\;\sum_{\substack{i\geq 0 \\[0.5mm] j> 0}} -
\sum_{\substack{i<0\\[0.5mm] j\leq 0}}\;\Biggr) (-1)^i
q^{\bigl(\ell+1-(i+2j)(k+2)\bigr)^2/(4(k+2))-(m+ik)^2/(4k)}\Biggr\}.
\notag 
\end{align}
Here $2r=m-\ell$ and $\eta$ is the Dedekin eta function:
$\eta(q)=q^{1/24}(q;q)_{\infty}$, with $(q;q)_n=\prod_{i=1}^n(1-q^i)$. 
The superscript $\Bup$ attached to $\hat{\chi}_{\ell,r}(q)$
indicates a bosonic or inclusion--exclusion form ---
a characteristic inherited from the form \eqref{WK} for the 
$\suh(2)_k$ characters.

Other derivations of $\hat{\chi}_{\ell,r}(q)$
have been given in the literature using a
free-field representation (one boson and a pair of ghosts
\cite{DQ} or three bosons \cite{Nem}) and the BRST construction.
This has led to bosonic expressions for the characters that are only
superficially different from \eqref{string}, and can easily
be proved to be equivalent by simple manipulations.

Shortly after the Kac--Petersen derivation, an intrinsically
different formula for the parafermion characters was obtained.
This is the famous Lepowsky--Primc expression \cite{LP} which 
yields the characters as a manifestly positive multiple-series
or fermionic (F) form (see also \cite{JMb}). Specifically,
\begin{equation}\label{lepo}
\hat{\chi}_{\ell,r}^{\Fup}(q)= q^{\Delta_{\ell}}
\sum_{\substack{n_1,\dots,n_{k-1}=0 \\
\frac{r+\ell}{k}+(C^{-1}n)_1\in\Z}}^{\infty}
\frac{q^{n C^{-1}(n-e_{\ell})}}
{(q;q)_{n_1}\cdots (q;q)_{n_{k-1}}},
\end{equation}
where $e_i$ denotes the $i$th standard unit vector in $\Z^{k-1}$ 
(with $e_0=e_k$ the zero vector),  $n=(n_1,\dots,n_{k-1})$ and $C$ 
denotes the A$_{k-1}$ Cartan matrix. (For more precise definitions of 
the employed notation, see Section~\ref{sec31}.)
The exponent $\Delta_{\ell}$ in the above is given by
\begin{equation}\label{Deltaell}
\Delta_{\ell}=-\frac{\ell^2}{4k}+\frac{(\ell+1)^2}{4(k+2)}-\frac{1}{12}.
\end{equation}

Yet a third expression for the parafermion characters can be
obtained from the so-called $\suh(2)_k$ spinon formula for
$\hat{\chi}_\ell(q)$ which was conjectured in \cite{BLS} and proved in
\cite{Ara,NYa,NYb}. Using \eqref{bfct} to extract the branching 
function from the spinon formula yields
\begin{equation}\label{spin}
\hat{\chi}_{\ell,r}^{\text{S}}(q)= (q;q)_{\infty}\, q^{\Delta_\ell}
\sum_{\substack{L,n_1,\cdots,n_{k-1}=0 \\ 
\frac{L+r+\ell}{k}+(C^{-1}n)_1\in\Z}}^{\infty}
\frac{q^{L(L+\ell+2r)/k+nC^{-1}(n-e_{\ell})}}
{(q;q)_L(q;q)_{L+\ell+2r}}
\prod_{i=1}^{k-1}\qbin{m_i+n_i}{n_i},
\end{equation}
where $\qbin{n}{m}$ is a $q$-binomial coefficient, and where
the integers $m_i$ can be computed from $L$ and the $n_i$ by
\begin{equation*}
m=C^{-1}\Bigl[(2L+r)e_1+e_{\ell}-2n\Bigr],
\end{equation*} 
with $m=(m_1,\dots,m_{k-1})$.

Finally, a fourth form of the parafermionic characters has been 
obtained in \cite{JM1}, relying on what can be called
`parafermionic representation theory.' 
The basic building block is the character of the parafermionic Verma 
module of relative charge $2t$,
given by
\begin{equation}\label{vermu}
\Vh_t(q)=\sum_{j=0}^{\infty}
\frac{q^{j+t}}{(q;q)_j(q;q)_{j+t}}.
\end{equation}
Combined with knowledge of the explicit structure of the singular 
vectors in the theory, the Verma character \eqref{vermu} 
leads to a second bosonic form of 
the characters as follows:
\begin{multline}\label{cara}
\hat{\chi}_{\ell,r}^{\Bupp}(q)
=q^{-\tfrac{1}{12}-\tfrac{(\ell+2r)^2}{4k}}
\sum_{j=-\infty}^{\infty}
q^{(k+2)\Bigl(j+\tfrac{\ell+1}{2(k+2)}\Bigr)^2}  \\
\times\Bigl\{\Vh_{r-(k+2)j}(q)-\Vh_{r+\ell+1+(k+2)j}(q)\Bigr\}.
\end{multline}

Viewed as formal $q$-series, the identities 
\begin{equation}\label{three}
\hat{\chi}_{\ell,r}^{\Bup}(q)=\hat{\chi}_{\ell,r}^{\text{S}}(q)
=\hat{\chi}_{\ell,r}^{\Fup}(q)
\end{equation}
have all been demonstrated in \cite{SW02}. To complete the formal proof of
the equality of all four parafermion expressions it 
remains to be shown that, for example,
\begin{equation}\label{bfB}
\hat{\chi}_{\ell,r}^{\Bup}(q)=\hat{\chi}_{\ell,r}^{\Bupp}(q).
\end{equation}
We fill this small gap in the literature in the appendix.

This concludes our review of character representations for ordinary
parafermions, and next we turn our attention to graded parafermions.

\subsection{Graded parafermion characters}
The present work is concerned with a demonstration 
--- at the level of $q$-series --- 
of the graded analogue of \eqref{three}, that is,
\begin{equation}\label{threegraded}
\chi_{\ell,r}^{\Bup}(q)=\chi_{\ell,r}^{\text{S}}(q)
=\chi_{\ell,r}^{\text{F}}(q).
\end{equation}
Here $\chi_{\ell,r}(q)$ is a graded parafermion characters
of relative charge $r$ and Dynkin label $\ell$, to be introduced below.
B, S and F again refer to bosonic, spinon and fermionic representations,
each of which arises from a
different algebraic formulation of the graded parafermion theory.

Graded parafermions were first introduced in \cite{CRS}.   
The underlying 
algebra is generated by `fermionic' parafermions $\psi_{1/2}$ and
$\psi_{1/2}^\dagger$ of dimension
$1-1/(4k)$ and central charge
\begin{equation}\label{cgpara}
c=-\frac{3}{2k+3}.
\end{equation}
The spectrum of the model has been determined in \cite{CRS}, and 
was shown to be equivalent to that of the coset
\begin{equation}\label{ospco}
\frac{\osph(1,2)_k}{\uh(1)},
\end{equation}
where $\osph(1,2)_k$ is the affine extension of the Lie 
superalgebra osp(1,2). The latter is a graded version of su(2) 
obtained by adding to the usual su(2) generators $J_{\pm}$ and $J_0$, 
the two fermionic generators $F_{\pm}$ such that $F_{\pm}^2=J_{\pm}$.

The equivalence between the graded parafermion theory and the 
coset \eqref{ospco} leads to the graded analogue of 
\eqref{bfct} and \eqref{bchihat}:
\begin{equation}\label{gbfct} 
\chi_{\ell}(x;q)=\sum_{m=1-k}^k
\chi_{\ell,{m-\ell}}(q) K_m(x;q),
\end{equation}
where $\chi_{\ell}(x;q)$ denotes the character of the
$\osph(1,2)_k$ integrable module of Dynkin label $\ell$
(with $0\leq \ell \leq k$), and 
$K_m(x;q)$ again denotes the $\uh(1)$ character of charge $m$.

A further analysis of the graded parafermion model
has been presented in \cite{JMc}, resulting in an
exact expression of the singular vectors and a description
--- in terms of jagged partitions ---
of the special nature of the spanning set of states. 
The resulting  expression 
for the Verma character of relative charge $t$ is given by
\begin{equation}\label{gradeV}
V_t(q)=\frac{1}{(q;q)_{\infty}}
\sum_{i,j=0}^{\infty}
\frac{q^{\binom{2j-2i-t}{2}+j}}{(q;q)_i(q;q)_j}.
\end{equation}
The associated character for the irreducible modules is an 
alternating sum over these Verma characters --- encoding the 
subtraction and addition of the embedded singular vectors --- and reads
\cite{BFJM}
\begin{multline}\label{carag}
\chi_{\ell,r}^{\Bup}(q)
=q^{-\tfrac{(\ell+r)^2}{4k}}
\sum_{j=-\infty}^{\infty}
q^{\tfrac{1}{2}(2k+3)\Bigl(j+\tfrac{2\ell+1}{2(2k+3)}\Bigr)^2}  \\
\times
\Bigl\{V_{r-(2k+3)j}(q)-
V_{r+2\ell+1+(2k+3)j}(q)\Bigr\}.
\end{multline} 
This is the bosonic formula for graded parafermions, 
one of the three central object of our work.  
Its underlying algebraic formulation is what could be dubbed 
the `graded parafermionic representation theory.'

In Section~\ref{secbf} we present an alternative, purely analytic
derivation of \eqref{carag} along the lines of the Kac--Peterson derivation 
of \eqref{string}, utilizing the branching rule
\eqref{gbfct} and the Weyl--Kac formula for the $\osph(1,2)_k$ characters.
Unlike the non-graded theory,
we have chosen to only work with a single bosonic representation,
dispensing of the need for $\Bupp$.

\medskip

The fermionic expression of a character reflects the construction 
of the corresponding module using a quasi-particle basis by a 
filling process subject to a generalized exclusion principle.
When the model is viewed directly from the point of view of the
parafermionic algebra, the modes of the parafermion $\psi_{1/2}$ 
are the natural choice for these quasi-particles.  
A descendant can thus be represented by the ordered sequence of 
its modes, and up to an overall sign this sequence forms a jagged 
partition \cite{JMc}. 
The generalized exclusion principle takes the form of certain
($k$-dependent) difference conditions on parts of the jagged partitions.
The generating function of the restricted jagged partitions yields, 
up to a minor adjustment, the fermionic expression for the characters 
of the irreducible modules:
\begin{multline}\label{fercar}
\chi_{\ell,r}^{\Fup}(q)=q^{h_{\ell}-c/24} \\
\times
\sum_{\substack{n_0,\dots,n_{k-1}=0 \\ 
\frac{r+2\ell+n_0}{2k}+(C^{-1}n)_1\in\Z}}^{\infty}
\frac{q^{-n_0(n_0+2\ell)/(4k)+\binom{n_0+1}{2}
+n C^{-1}(n-n_0 e_1-e_{\ell})}}
{(q;q)_{n_0}\cdots (q;q)_{n_{k-1}}}.
\end{multline}
Here the exponent $h_{\ell}$ is given by the $r=0$ instance
of \eqref{hlr} of Section~\ref{sec21}.

\medskip

Finally, we describe a third form for graded parafermion characters, 
referred to as a spinon formula. 
It is inherited from the following $\osph(1,2)_k$ analogue
of the  $\suh(2)_k$ spinon formula:
\begin{multline}\label{gspin}
\chi_{\ell}(x;q)=
q^{h_{\ell}-c/24+\ell^2/(4k)-1/24}
\sum_{\substack{L_{+},L_{-},n_0,\cdots,n_{k-1}=0 \\ 
\frac{L_{+}+L_{-}+n_0+\ell}{2k}+(C^{-1}n)_1\in\Z}}^{\infty}
x^{-(L_{+}-L_{-})/2} \\ 
\times
\frac{q^{[(L_{+}+L_{-})^2-(n_0+\ell)^2]/(4k) 
+\binom{n_0+1}{2}+nC^{-1}(n-n_0e_1-e_{\ell})}}
{(q;q)_{L_{+}}(q;q)_{L_{-}}(q;q)_{n_0}}
\prod_{i=1}^{k-1}\qbin{n_i+m_i}{n_i}, 
\end{multline}
where
\begin{equation*}
m=C^{-1}\Bigl[(L_{+}+L_{-}+n_0)e_1+e_{\ell}-2n \Bigr].
\end{equation*}

The formula \eqref{gspin} for the $\osph(1,2)_k$ character
was initially conjectured using the analogy with the non-graded
theory.
In the non-graded context, the term spinon refers to the
two components of the basic $\ell=1$ $\suh(2)_k$ WZW primary field.
In the present case however,
the doublet is traded by a triplet, the third component
manifesting itself in the presence of the $n_0$ mode in \eqref{gspin}.
Its derivation from a spinon-type basis of states remains to be worked out.

The  expression for the parafermionic character that results
from \eqref{gspin} and \eqref{gbfct} is, with
$L_{+}=L+r+\ell$ and $L_{-}=L$,
\begin{multline}\label{gparaspi}
\chi_{\ell,r}^{\Sup}  (q)=(q;q)_{\infty}\, q^{h_\ell-c/24}\!
\sum_{\substack{L,n_0,\cdots,n_{k-1}=0 \\ 
\frac{2L+r+2\ell+n_0}{2k}+(C^{-1}n)_1\in\Z}}^{\infty} \!\!
q^{L(L+r+\ell)/k-n_0(n_0+2\ell)/(4k)+\binom{n_0+1}{2} }\\
\times \frac{q^{nC^{-1}(n-n_0e_1-e_{\ell})}}
{(q;q)_L (q;q)_{L+r+\ell} (q;q)_{n_0}}
\prod_{i=1}^{k-1}\qbin{n_i+m_i}{n_i},
\end{multline}
where
\begin{equation*}
m=C^{-1}\Bigl[(2L+r+\ell+n_0)e_1+e_{\ell}-2n \Bigr].
\end{equation*}
Our proof of \eqref{gparaspi} in Section~\ref{secBSF} of course
also confirms the validity of \eqref{gspin}.

\subsection{Outline of the proof of \eqref{threegraded}}
The proof of the equivalence of the three character representations
\eqref{carag}, \eqref{fercar} and \eqref{gparaspi} is presented in Section~\ref{secBSF},
and forms the core of this paper.
Due to its very technical nature, the proof
may --- certainly at first reading --- not be very insightful.
It thus seems important to point out the main ideas and
underlying physics. 

Given our much better understanding of the non-graded parafermions,
an obvious approach to establishing \eqref{threegraded} is to
connect with the non-graded character theory.
To this end we use a little trick and write the coset \eqref{ospco}
as
\begin{equation}\label{cosetdec}
\frac{\osph(1,2)_k}{\uh(1)} = 
\frac{\osph(1,2)_k}{\suh(2)_k}\times \frac{\suh(2)_k }{\uh(1)},
\end{equation}
where the second coset on the right is that of ordinary parafermions,
see \eqref{cos}.
In \cite{FY,KW} it was shown that 
\begin{equation}\label{Mk}
\frac{\osph(1,2)_k}{\suh(2)_k}\simeq \Mc(k+2,2k+3),
\end{equation}
where the right-hand side denotes the non-unitary
minimal model of central charge
\begin{equation*}
c^{(k+2,2k+3)}=1-\frac{6(k+1)^2}{(k+2)(2k+3)}.
\end{equation*}
(The reader may readily check that
$-\frac{3}{2k+3}=c^{(k+2,2k+3)}+\frac{2(k-1)}{k+2}$.  )

Now let $\tilde{\chi}^{(p,p')}_{r,s}(q)$ be the
usual $\Mc(p,p')$ Virasoro character \cite{CFT}
\begin{equation}\label{Virchar}
\tilde{\chi}^{(p,p')}_{r,s}(q)
=\frac{1}{\eta(q)}\sum_{\sigma\in\{\pm 1\}}
\sum_{j=-\infty}^{\infty} \sigma
q^{pp'\Bigl(j+\tfrac{p'r-
\sigma p s}{2p p'}\Bigr)^2}.
\end{equation}
Then, by \eqref{Mk} and 
\begin{equation*}
\osph(1,2)_k= 
\frac{\osph(1,2)_k}{\suh(2)_k}\times \suh(2)_k,
\end{equation*}
we have
\begin{equation}\label{noproof}
\chi_{\ell}(x;q)=\sum_{i=0}^k 
\tilde{\chi}^{(k+2,2k+3)}_{i+1,2\ell+1}(q)
\hat{\chi}_i(x;q).
\end{equation}
By \eqref{bfct}, \eqref{bchihat} (plus
$\hat{b}_{\ell,m}(q)=0$ when $m-\ell$ is odd) and
\eqref{gbfct} it thus follows that the decomposition
\eqref{cosetdec} gives rise to
\begin{equation}\label{formula}
\chi_{\ell,r-\ell}(q)=
\sum_{\substack{i=0 \\ i\equiv r\ppmod{2}}}^k
\tilde{\chi}_{i+1,2\ell+1}^{(k+2,2k+3)}(q)\hat{\chi}_{i,\frac{r-i}{2}}(q).
\end{equation}
It is this formula that provides the necessary handle on 
\eqref{threegraded}.

Indeed, both the Virasoro and (ordinary) parafermion
character on the right admit an interpretation 
in terms of the same combinatorial objects: the 
one-dimensional configuration sums
$X_{r,s}^{(p,p')}(L,b;q)$ of the
Andrews--Baxter--Forrester models.
This leads to (see Section~\ref{sec0})
\begin{multline}\label{formula2}
 \chi_{\ell,r-\ell}(q)=
q^{h_{\ell}-{(r-\ell)(r+\ell)/(4k)}-c/24}\\
\times (q;q)_{\infty}
\sum_{\substack{n=0 \\ n\equiv r+\ell\ppmod{2}}}^{\infty}
\frac{q^{n^2/2}}{(q;q)_n} 
\sum_{L=0}^{\infty}\frac{q^{(2L+r-\ell)/2}}
{(q;q)_L(q;q)_{L+r}} \\ \times
\sum_{\substack{i=0 \\ i\equiv r \ppmod{2}}}^k
X_{1,i+1}^{(1,k+2)}(n,\ell+1;q^{-1})X_{1,i+1}^{(1,k+2)}(2L+r,1;q).
\end{multline}

Our proof of \eqref{threegraded} basically amounts to showing
that \eqref{formula2} is compatible with both 
\eqref{carag} and \eqref{gparaspi}. 
In the case of \eqref{gparaspi} this requires a combinatorial
(subtractionless or fermionic)
method for computing the sum over products of 
one-dimensional configuration sums, whereas in
the case of \eqref{carag} our approach will be
analytic in nature.
Once the compatibility of \eqref{formula2} with
\eqref{carag} and \eqref{gparaspi} is established,
we have achieved the
desired equality $\chi_{\ell,r}^{\Bup}(q)=
\chi_{\ell,r}^{\Sup}(q)$.

Showing the remaining
\begin{equation*}
\chi_{\ell,r}^{\Sup}(q) = \chi_{\ell,r}^{\Fup}(q)
\end{equation*}
does not rely on \eqref{formula2}, but
closely follows the approach of \cite{SW02} for
proving 
\begin{equation*}
\hat{\chi}_{\ell,r}^{\Sup}(q) = \hat{\chi}_{\ell,r}^{\Fup}(q)
\end{equation*}
in the case of non-graded parafermions.

\bigskip

Given that the bosonic formula \eqref{carag} significantly simplifies 
in the large $k$ limit, it is of interest to look for  
an entirely analytic derivation of the equivalence of 
$\chi_{\ell,r}^{\Bup}(q)$ and $\chi_{\ell,r}^{\Fup}(q)$ 
in this limit.
This will be the content of Section~\ref{sec4}.

\section{Graded parafermions and $\osph(1,2)_k$ characters}

\subsection{Graded parafermions}\label{sec21}
The graded parafermion conformal theory is defined by  an algebra 
that generalizes the usual $\Z_k$ parafermionic algebra 
(i.e., $\psi_n\times \psi_m\sim \psi_{n+m}$ with $\psi_k\sim I$), 
by the addition of a $\Z_2$ grading $\psi_{1/2}\times\psi_{1/2}\sim\psi_1$ 
\cite{CRS}. The new parafermion $\psi_{1/2}$ has conformal dimension
$1-1/(4k)$. Associativity of the operator product algebra fixes the 
central charge as in \eqref{cgpara}
The parafermionic primary fields
$\phi_\ell$ are labeled by an integer $\ell$ such that
$0\leq\ell\leq k$.
If we denote the corresponding highest-weight state by $|\phi_\ell\Ra$,
then modules over $|\phi_\ell\Ra$ can be decomposed into a finite sum 
of modules with specific charge.
The latter is normalized by setting the charge of $\psi_{1/2}$ equal to one,
and is in fact defined modulo $2k$ only, since $(\psi_{1/2})^{2k}\sim I$.
The highest-weight state of relative charge $r$
in the highest-weight module labeled by $\ell$,
has dimension
\begin{equation}\label{hlr}
h_{\ell}^{(r)}=\frac{\ell(2k-3\ell)}{4k(2k+3)} -
\frac{r(2\ell+r)}{4k} +\max\{0,\lfloor(r+1)/2\rfloor\},
\end{equation}
where $h_{\ell}^{(0)}$ will be simply denoted as $h_{\ell}$.

In terms of the coset description \eqref{ospco}, the label $\ell$ 
is the finite Dynkin label
of $\osph(1,2)_k$ and $r$ is related to the u(1) charge $m$ by
\begin{equation*}
r=m-\ell.
\end{equation*}
With this identification, the dimension $h_{\ell}^{(r)}$ can be rewritten as
\begin{equation*}
h_{\ell}^{(m-\ell)}=
\frac{\ell(\ell+1)}{2(2k+3)}-\frac{m^2}{4k}+
\max\{0,\lfloor(m-\ell+1)/2\rfloor\},
\end{equation*}
which can be recognized as the difference (modulo integers)
between the dimension of
the $\osph(1,2)_k$ primary field with spin $\ell/2$ and that of the 
$\uh(1)$ field of charge $m$.

In order to specify the highest-weight conditions, let us recall that 
when acting on a generic state of charge $t$, denoted by $|t\Ra$, 
the parafermionic modes $B_n$ are defined as
\begin{align*}
\psi_{\frac{1}{2}}(z)|t \Ra
&=\sum_{m=-\infty}^{\infty}
z^{-t/2k-m-1}B_{(1+2t)/4k+m}|t\Ra, \\
\psi_{\frac{1}{2}}^{\dagger}(z)|t\Ra&=\sum_{m=-\infty}^{\infty}
z^{t /2k-m-1} B^{\dagger}_{(1-2t)/4k+m}|t\Ra.
\end{align*}
In the following, we use the more compact notation
\begin{equation*}
\B_n|t\Ra:=B_{n+(1+2t)/4k}|t\Ra, \qquad
\B_n^{\dagger}|t\Ra:= B_{n+(1-2t)/4k}^{\dagger}|t\Ra.
\end{equation*}
Then the defining relations for the highest-weight 
states $|\phi_\ell\Ra$ are
\begin{equation*}
\B_n |\phi_{\ell}\Ra=0=\B^{\dagger}_{n+1}|\phi_{\ell}\Ra,
\qquad n\geq 0.
\end{equation*}
The dimension $h_\ell$ of the highest-weight state $|\phi_\ell \Ra$ 
follows directly from these constraints together with the generalized 
commutation relations induced by the operator product algebra. 
The dimension $h_{\ell}^{(r)}$, in turn, is $h_{\ell}$ plus the dimension
of $(\B_{-1})^r$  for $r>0$ or $(\B_0^\dagger)^{-r}$ for $r<0$.

\subsection{The Verma characters}
The free module over the highest-weight state  $|\phi_\ell\Ra$ can be 
decomposed into a direct sum of modules with fixed relative charge $r$.
The states of the module of charge $r\geq 0$ are of the form
\begin{equation}\label{lamu} 
\B_{-\la_1}\B_{-\la_2}\cdots 
\B_{-\la_p}\B^{\dagger}_{-\mu_1}\B^{\dagger}_{-\mu_2}\cdots
\B^{\dagger}_{-\mu_{p'}} |\phi_\ell\Ra,
\end{equation}
where 
\begin{equation}\label{pppr}
p-p'=r
\end{equation}
and
\begin{subequations}\label{jag}
\begin{align}
\la_i &\geq \la_{i+1}-1, & \la_i &\geq \la_{i+2}, &\la_p &\geq 1 \\ 
\mu_i &\geq \mu_{i+1}-1, & \mu_i &\geq \mu_{i+2}, &\mu_i &\geq 0.
\end{align}
\end{subequations}
Adopting the terminology of \cite{JMc} we refer to the
sequence $\la=(\la_1,\cdots,\la_p)$ as a jagged partition of length 
$p$ and weight $\la_1+\cdots+\la_p$.
Clearly, the sequence $\mu=(\mu_1,\dots,\mu_{p'})$ then
corresponds to a jagged partition of length at most $p'$.

In order to compute the character $V_r(q)$ of the free module 
of relative charge $r$ --- Verma character for short --- we 
denote by $J_{p,k}$ and $\bar{J}_{p',k}=\sum_{p=0}^{p'} J_{p,k}$
the number of jagged partitions of length $p$ and the
number of jagged partitions of length at most $p'$.
For example, the four jagged partitions of length $4$ and weight $5$
are $(3,1,0,1)$, $(2,2,0,1)$, $(2,1,1,1)$ and $(1,2,1,1)$, so that
$J_{4,5}=4$.
We further introduce the generating functions
\begin{align*}
J(z;q)&=\sum_{p=0}^{\infty} J_p(q) z^p
=\sum_{p,k=0}^{\infty} J_{p,k}z^p q^k \\
\bar{J}(z;q)&=\sum_{p=0}^{\infty} \bar{J}_{p}(q)z^p
=\sum_{p,k=0}^{\infty} \bar{J}_{p,k}z^p q^k \\
\end{align*}
so that $J_p(q)$ and $\bar{J}_p(q)$ are the generating
functions of jagged partitions of length $p$ and of length
at most $p$, respectively.
Since $J_{p,k}=\bar{J}_{p,k}-\bar{J}_{p-1,k}$ we immediately
infer that
\begin{equation}\label{JJbar}
J(z;q)=(1-z)\bar{J}(z;q).
\end{equation}

According to \eqref{pppr} and \eqref{jag} the Verma character
is expressed in terms of the above generating functions as
\begin{equation}\label{VJJ}
V_r(q)=\sum_{\substack{p,p'=0 \\ p-p'=r}}^{\infty}
J_p(q)\bar{J}_{p'}(q).
\end{equation}
To obtain a closed form expression for $V_r(q)$ we need some standard
$q$-series notation.
Let $(a;q)_0=1$,
\begin{equation}\label{qfac}
(a;q)_n=\prod_{k=0}^{n-1}(1-aq^k)
\end{equation}
for $n$ a positive integer and
\begin{equation*}
(a;q)_{\infty}=\prod_{k=0}^{\infty}(1-aq^k), \qquad \abs{q}<1.
\end{equation*}
Using that $(a;q)_n=(a;q)_{\infty}/(aq^n;q)_{\infty}$ one can
extend the definition \eqref{qfac} to all integers $n$.
In particular this implies that $1/(q;q)_{-n}=0$ for
$n$ a positive integer.

Returning to our calculation of the Verma character 
we recall that \cite{BFJM}
\begin{equation}\label{Jprod}
J(z;q)=\frac{(-zq;q)_{\infty}}{(z^2q;q)_{\infty}}.
\end{equation}
By Euler's two formulae \cite[Eqs. (II.1) and (II.2)]{GR}
\begin{subequations}
\begin{align}\label{eulera}
(z;q)_{\infty}&=\sum_{j=0}^{\infty}
\frac{(-z)^j q^{\binom{j}{2}}}{(q;q)_j}, \\
\frac{1}{(z;q)_{\infty}}&=\sum_{j=0}^{\infty}\frac{z^j}{(q;q)_j},
\label{euler}
\end{align}
\end{subequations}
it thus follows that
\begin{equation}\label{Jm}
J_p(q)=\sum_{j=0}^{\infty}
\frac{q^{\binom{p-2j+1}{2}+j}}{(q;q)_{p-2j}(q;q)_j}.
\end{equation}
{}From \eqref{JJbar} and \eqref{Jprod} we find that
\begin{equation*}
\bar{J}(z;q)=\frac{1}{1-z}\frac{(-zq;q)_{\infty}}{(z^2q;q)_{\infty}}
=\frac{(-z;q)_{\infty}}{(z^2;q)_{\infty}}.
\end{equation*}
Again using the Euler formulae this implies a companion to \eqref{Jm}
as follows:
\begin{equation}\label{Jbar}
\bar{J}_{p'}(q)
=\sum_{j=0}^{\infty} \frac{q^{\binom{p'-2j}{2}}}{(q;q)_{p'-2j}(q;q)_j}.
\end{equation}
An alternative derivation of \eqref{Jbar} follows from 
Lemma 17 and Equations (56) and (58) of \cite{FJM}.

Substituting \eqref{Jm} and \eqref{Jbar} in \eqref{VJJ} 
results in
\begin{equation*}
V_r(q)=\sum_{p,i,j=0}^{\infty}
\frac{q^{\binom{p-2j+1}{2}+\binom{p-r-2i}{2}+j}}
{(q;q)_{p-r-2i}(q;q)_i(q;q)_{p-2j}(q;q)_j}.
\end{equation*}
Shifting $p\to p+2j$ and then summing over $p$ using
the Durfee rectangle identity \cite[Eq. (II.8)]{GR}
\begin{equation}\label{Durfee}
\sum_{k=0}^{\infty}\frac{q^{k(k+a)}}{(q;q)_k(q;q)_{k+a}}
=\frac{1}{(q;q)_{\infty}}, \qquad a\in\Z,
\end{equation}
yields the expression for the Verma 
character of relative charge $r$ given in \eqref{gradeV}.
This should be compared with the quadruple sum representation for 
$V_r(q)$ of~\cite{BFJM}.

The Verma character is the character of a generic
highest-weight module not containing singular vectors.
Modules over $|\phi_\ell\Ra$ are highly reducible. 
A closed-form expression can be obtained for the singular vectors, 
from which their dimension and charge are easily read off. 
The irreducible modules are thus obtained by factoring these vectors 
using an exclusion-inclusion process.
As a result, the characters of an irreducible module of 
relative charge $r$ are
expressed as the alternating sum of Verma characters given in \eqref{carag}.

\subsection{The $\osph(1,2)_k/\uh(1)$ branching functions}\label{secbf}

In this section we rederive \eqref{carag}
through a direct computation of the branching functions \eqref{ospco}.
This derivation is analogous to 
the one presented in the appendix
for \eqref{cara}.
   
The character of the $\osph(1,2)_k$ integrable module
indexed by the integer Dynkin label $0\leq \ell \leq k$ is 
\cite[App. A]{ER}\footnote{
Recall that for osp(1,2), the dual Coxeter number is
$3/2$ and the Weyl vector is $\rho= \omega_1/2$ where
$\omega_1$ is the su(2) fundamental weight.}
\begin{equation}\label{caosp}
\chi_{\ell}(x;q) =
\frac{\Theta^{(k+3/2)}_{\ell+1/2}(x;q)-
\Theta^{(k+3/2)}_{-\ell-1/2}(x;q)}
{\Theta^{(3/2)}_{1/2}(x;q)-\Theta^{(3/2)}_{-1/2}(x;q)}.
\end{equation}
Here $\Theta_m^{(k)}$ is the theta function
\begin{equation} 
\Theta_m^{(k)}(x;q)=\sum_{n=-\infty}^{\infty} 
q^{k\bigl(n+\tfrac{m}{2k}\bigr)^2} 
x^{-k\bigl(n+\tfrac{m}{2k}\bigr)}.
\label{Theta}
\end{equation}
The $\uh(1)$ character for a module with u(1) charge $m$ is
\begin{equation*}
K_m(x;q)=\frac{\Theta_m^{(k)}(x;q)}{\eta(q)}.
\end{equation*}
This form manifests the symmetry
\begin{equation*}
K_m(x;q)=K_{m+2k}(x;q).
\end{equation*}

The claim that the graded parafermion characters arise as the
branching functions of the $\osph(1,2)_k/\uh(1)$ coset
translates into the identity \eqref{gbfct}.
The aim of this section is to show that this
is in accordance with the bosonic representation of 
$\chi_{\ell,r}(q)$ as given in \eqref{carag}.

In passing we note that the $\osph(1,2)_k$
string functions are defined by the decomposition
\begin{equation*}
\chi_{\ell}(x;q)=\sum_{m=1-k}^k c_m^{\ell}(q)\Theta_m^{(k)}(x;q),
\end{equation*}
so that \eqref{gbfct} is equivalent to
\begin{equation*}
\chi_{\ell,{m-\ell}}(q)=\eta(q) c_m^{\ell}(q).
\end{equation*}

Now, to obtain a formal expansion for the character 
$\chi_{\ell}(x;q)$ 
we use the quintuple product identity \cite[Ex. 5.6]{GR}
\begin{equation*}
\sum_{n=-\infty}^{\infty}q^{n(3n-1)/2}x^{3n}(1-xq^n)=
(x,q/x,q;q)_{\infty}(qx^2,q/x^2;q^2)_{\infty}, \quad x\neq 0,
\end{equation*}
where $(a_1,a_2,\dots,a_k;q)_n=(a_1;q)_n(a_2;q)_n\cdots (a_k;q)_n$,
to rewrite the denominator of \eqref{caosp} (where we have set $x=y^2$)
as
\begin{align*}
\Theta^{(3/2)}_{1/2}(y^2;q)-\Theta^{(3/2)}_{-1/2}(y^2;q)
&=q^{1/24}\,y^{-1/2}\sum_{n=-\infty}^{\infty}
q^{n(3n-1)/2}y^{3n}(1-yq^n) \\
&=q^{1/24}\,y^{-1/2}(y,q/y,q;q)_{\infty}
(qy^2,q/y^2;q^2)_{\infty} \\
&=q^{1/24}\,y^{-1/2}\frac{(y^2,q/y^2,q;q)_{\infty}}
{(-y,-q/y;q)_{\infty}}.
\end{align*}
By a double use of \eqref{euler} we obtain
\begin{equation}\label{keyexp}
\frac{1}{(x,q/x;q)_{\infty}}=\sum_{i,j=0}^{\infty}
\frac{q^j x^{i-j}}{(q;q)_i(q;q)_j}, \qquad  \abs{q}<\abs{x}<1.
\end{equation}
This and the Jacobi triple product identity \cite[Eq. (II.28)]{GR}
\begin{equation}\label{jaco}
\sum_{n=-\infty}^{\infty}q^{\binom{n}{2}}x^n=
(-x,-q/x,q;q)_{\infty}, \qquad x\neq 0,
\end{equation}
imply that
\begin{align*}
\frac{(-y,-q/y;q)_{\infty}}{(y^2,q/y^2,q;q)_{\infty}}
&=\frac{1}{(q;q)_{\infty}^2}
\sum_{n=-\infty}^{\infty}\sum_{i,j=0}^{\infty}
\frac{y^{n+2i-2j}q^{\binom{n}{2}+j}}{(q;q)_i(q;q)_j} \\
&=\frac{1}{(q;q)_{\infty}^2}
\sum_{n=-\infty}^{\infty}\sum_{i,j=0}^{\infty}
\frac{y^{-n} q^{\binom{2j-2i-n}{2}+j}}{(q;q)_i(q;q)_j} \\
&=\frac{1}{(q;q)_{\infty}}\sum_{n=-\infty}^{\infty} y^{-n} V_n(q),
\qquad \qquad \abs{q}<\abs{y}^2<1,
\end{align*}
where the last equality follows from \eqref{gradeV}.
We thus arrive at the following
expansion of the denominator of \eqref{caosp}
in terms of Verma characters:
\begin{equation*}
\frac{1}{\displaystyle\sum_{\sigma\in\{\pm 1\}}
\Theta^{(3/2)}_{\sigma/2}(x;q)}
=\frac{x^{1/4}}{\eta(q)}\sum_{n=-\infty}^{\infty} x^{-n/2} V_n(q).
\end{equation*}

Multiplying the above expression by the numerator of \eqref{caosp} yields
\begin{multline*}
\chi_{\ell}(x;q) 
=\frac{1}{\eta(q)}
\sum_{\sigma\in\{\pm 1\}}\sigma \sum_{j,n=-\infty}^{\infty}
x^{-n/2+j(2k+3)/2+\sigma(\ell+1/2)/2+1/4} \\
\times
q^{\tfrac{1}{2}(2k+3)\Bigl(\sigma j+\tfrac{2\ell+1}{2(2k+3)}\Bigr)^2} V_n(q).
\end{multline*}
We now arrange the $x$ dependence into sums
over the $\uh(1)$ characters.
Shifting $n\to n-(2k+3)j-\sigma(\ell+1/2)+1/2$ and replacing
$j$ by $\sigma j$ this is achieved as
\begin{align*}
\chi_{\ell}(x;q)&=\frac{1}{\eta(q)}
\sum_{n=-\infty}^{\infty} x^{-n/2}
\sum_{j=-\infty}^{\infty}
q^{\tfrac{1}{2}(2k+3)\Bigl(j+\tfrac{2\ell+1}{2(2k+3)}\Bigr)^2}  
\\ & \qquad \qquad \qquad \qquad \qquad \times
\sum_{\sigma\in\{\pm 1\}}\sigma
V_{n-\sigma (2k+3)j-\sigma(\ell+1/2)+1/2}(q) \\
&=\frac{1}{\eta(q)}
\sum_{n=-\infty}^{\infty} x^{-n/2} q^{n^2/4k}
\chi_{\ell,n-\ell}(q).
\end{align*}
with  $\chi_{\ell,r}(q)$ given by \eqref{carag}.
To complete the decomposition of the 
$\osph(1,2)_k$ character, we set
$n=2kj+m$ with $1-k\leq m\leq k$ and use the symmetry 
$\chi_{\ell,m}(q)=\chi_{\ell,m+2kj}(q)$.
This gives 
\begin{align*}
\chi_{\ell}(x;q)&=\sum_{m=1-k}^k
\chi_{\ell,m-\ell}(q)\sum_{j=-\infty}^{\infty}
\frac{q^{k(j+m/2k)^2} x^{-k(j+m/2k)}}{\eta(q)}\\
&=\sum_{m=1-k}^k \chi_{\ell,m-\ell}(q) K_m(x;q)
\end{align*}
in agreement with \eqref{carag} and \eqref{gbfct}.

\subsection{Fermionic parafermion characters}\label{sec31}
Fermionic forms for the characters
of the graded parafermion models were first constructed in \cite{BFJM}.
Correcting a minor misprint, the result of \cite{BFJM} states that
\begin{multline}\label{finitek}
\chi_{\ell,r}^{\Fup}(q)= q^{h_{\ell}-c/24} \\
\times\!\!\!\!
\sum_{\substack{n_0,\dots,n_{k-1}=0 \\ n_0+2N\equiv r\ppmod{2k}}}^{\infty}
\!\!\!
\frac{q^{\binom{n_0+1}{2}+N_1^2+\cdots+N_{k-1}^2+N_{k-\ell+1}+\cdots+N_{k-1}
-(n_0+2N)(n_0+2N+2\ell)/(4k)}}
{(q;q)_{n_0}\cdots (q;q)_{n_{k-1}}},
\end{multline}
with $N_i=n_i+\cdots+n_{k-1}$, $N=N_1+\cdots+N_{k-1}$ and $N_k=N_{k+1}=0$.

The above fermionic form encodes the combinatorics of 
the graded parafermionic quasi-particle basis constructed in \cite{JMc}. 
This basis is described by states of the form
\begin{equation*}
\B_{-\la_1}\B_{-\la_2}\cdots \B_{-\la_p}|\phi_\ell\Ra
\end{equation*}
for $p=mk+r$. Here the $\la_i$ are again subject to \eqref{jag} with 
$\la_p\geq 1$, but --- in the case of strictly positive $k$ --- 
are further constrained by
\begin{equation*}
\la_j \geq \la_{j+2k-1}+1 \qquad \text{or} \qquad
\la_j=\la_{j+1}-1=\la_{j+2k-2}+1=\la_{j+2k-1},
\end{equation*}
for all $j\leq p-2k+1$.

In the derivation of \eqref{finitek} it is understood that $r\geq 0$ 
(that is, we act with $r$ modes modulo $k$ on the highest-weight state). 
We shall later show, however, that the expression \eqref{finitek}
obeys the formal symmetry relation
\begin{equation}\label{sym}
\chi_{\ell,r}(q)=\chi_{\ell,-r-2\ell}(q).
\end{equation}
This will allow us to interpret $\chi_{\ell,r}(q)$ for all integers $r$.

To connect with the fermionic form for the characters stated
in \eqref{fercar} we first need to replace
$n_i\to n_{k-i}$ for $1\leq i\leq k-1$ and use the
symmetry \eqref{sym}.
Then introducing the vector notation
\begin{equation*}
wAv=\sum_{i,j=1}^{k-1}w_i A_{ij}v_j
\qquad\text{and}\qquad (Av)_i=\sum_{j=1}^{k-1}A_{ij}v_j
\end{equation*}
for $w,v\in\Z^{k-1}$ and $A$ a square matrix of dimension $k-1$,
we get \eqref{fercar} with
$C$ the A$_{k-1}$ Cartan matrix:
\begin{equation}
C^{-1}_{ij}=\min\{i,j\}-\frac{ij}{k},
\end{equation}
$e_i$ the $i$th standard unit vector in $\Z^{k-1}$
(with $e_0=e_k$ the zero vector) and
$n=(n_1,\dots,n_{k-1})$.

We note that since
the entries of $C^{-1}$ are integer multiples of  
$1/k$, the restriction imposed on the sum over the $n_i$
in \eqref{fercar} implies that the summand vanishes unless
$n_0-r$ is even.

\section{Proof of $\Bup=\Sup=\Fup$ for graded parafermions}\label{secBSF}

This section, in which we prove the equivalence of the
bosonic, fermionic and spinon forms of the graded parafermion
characters, forms the core of the paper.
\subsection{Main results}
Normalising the characters in \eqref{threegraded} we set out
to prove the following theorem.
Let 
\begin{equation}\label{qbinom}
\qbin{n}{k}_q=\qbin{n}{k}=
\begin{cases}\displaystyle
\frac{(q;q)_n}{(q;q)_k(q;q)_{n-k}}\quad  &
k\in\{0,1,\dots,n\}, \\[3mm]
0 & \text{otherwise}
\end{cases}
\end{equation}
be a $q$-binomial coefficient,
and assume the vector notation introduced in Section~\ref{sec31}.
\begin{theorem}\label{thm1}
Let $k$ be a positive integer.
For $\ell\in\{0,\dots,k\}$ and $r\in\Z$ there holds
\begin{align}\label{BF}
\sum_{j=-\infty}^{\infty} &
q^{j((2k+3)j+2\ell+1)/2}
\Bigl\{V_{r-\ell-(2k+3)j}(q)-
V_{r+\ell+1+(2k+3)j}(q)\Bigr\} \\
&=(q;q)_{\infty}
\sum_{\substack{L,n_0,\dots,n_{k-1}=0 \\
\frac{2L+r+\ell+n_0}{2k}+(C^{-1}n)_1\in\Z}}^{\infty}
\!\!
q^{L(L+r)/k+(r-\ell-n_0)(r+\ell+n_0)/(4k)+
\binom{n_0+1}{2}} \notag \\
&\qquad\qquad\qquad\qquad\qquad\qquad\times
\frac{q^{n C^{-1}(n-n_0 e_1-e_{\ell})}}
{(q;q)_L(q;q)_{L+r}(q;q)_{n_0}} 
\prod_{i=1}^{k-1}\qbin{n_i+m_i}{n_i}
\notag  \\[2mm]
&=\sum_{\substack{n_0,\dots,n_{k-1}=0 \\
\frac{r+\ell+n_0}{2k}+(C^{-1}n)_1\in\Z}}^{\infty}
\frac{q^{(r-\ell-n_0)(r+\ell+n_0)/(4k)+\binom{n_0+1}{2}+n C^{-1}(n-n_0
e_1-e_{\ell})}}
{(q;q)_{n_0}\cdots (q;q)_{n_{k-1}}}. \notag
\end{align}
Here the $m_i$ in the expression after the first equality
follow from
\begin{equation}\label{mn}
m=C^{-1}\Bigl[(2L+r+n_0)e_1+e_{\ell}-2n\Bigr].
\end{equation}
\end{theorem}
Similar to our earlier comment we remark that
the summands of the two multiple sums on the right vanish 
when $n_0+r+\ell\not\equiv 0\pmod{2}$.

\medskip

For the sake of brevity we write the statement of Theorem~\ref{thm1}
as
\begin{equation}\label{LR}
\psi^{\Bup}_{\ell,r}(q)=
\psi^{\Sup}_{\ell,r}(q)=
\psi^{\Fup}_{\ell,r}(q).
\end{equation}
Comparison with \eqref{carag}, \eqref{fercar} 
and \eqref{gparaspi} shows that
\begin{equation}\label{psichi}
\chi_{\ell,r-\ell}(q)=
q^{h_{\ell}-{(r-\ell)(r+\ell)/(4k)}-c/24}
\psi_{\ell,r}(q)
\end{equation}
so that \eqref{LR} implies \eqref{threegraded}.

The remainder of this section will be devoted to a proof of
Theorem~\ref{thm1}, partially by analytic and partially by combinatorial
means. Because of the length and complexity of the proof,
we will state the main intermediate results leading to the theorem 
in the form of three propositions.
Section~\ref{sec0} will provide some further insight
into the origin of the first proposition (which
essentially is a restatement of \eqref{formula2}).
Then, in Sections~\ref{secprop1}--\ref{secprop3}, 
each of the propositions will be proved.

\medskip

Before we can state our first proposition
we need to recall the definition of the one-dimensional
configuration sums of the Andrews--Baxter--Forrester model,
given by \cite{ABF84,FB85}
\begin{multline}\label{ABF}
X_{r,s}^{(p,p')}(L,b;q)
=\sum_{j=-\infty}^{\infty} 
\biggl\{q^{j(pp'j+p'r-ps)}\qbin{L}{(L+s-b)/2-p'j} \\
- q^{(pj+r)(p'j+s)}\qbin{L}{(L-s-b)/2-p'j}\biggr\}.
\end{multline}
Here $p,p',r,s,b$ and $L$ are integers such that
$1\leq p\leq p'$, 
$1\leq b,s\leq p'-1$, $0\leq r\leq p$,
and $L+s+b\equiv 0\pmod{2}$.

Our first proposition allows for the rewriting of the left-hand side of
\eqref{LR} in terms of one-dimensional configuration sums,
and may be recognized as a normalized version of
\eqref{formula2}.
\begin{proposition}\label{prop1}
For $r$ an integer and $\ell\in\{0,\dots,k\}$ there holds
\begin{multline*}
\psi_{\ell,r}^{\Bup}(q)=(q;q)_{\infty} 
\sum_{\substack{n=0 \\ n\equiv r+\ell\ppmod{2}}}^{\infty}
\frac{q^{n^2/2}}{(q;q)_n} 
\sum_{L=0}^{\infty}\frac{q^{(2L+r-\ell)/2}}
{(q;q)_L(q;q)_{L+r}} \\ \times
\sum_{\substack{i=0 \\ i\equiv r \ppmod{2}}}^k
X_{1,i+1}^{(1,k+2)}(n,\ell+1;q^{-1})X_{1,i+1}^{(1,k+2)}(2L+r,1;q).
\end{multline*}
\end{proposition}
Since $1/(q;q)_{-n}=0$ for $n$ a positive integer, the lower bound
in the sum over $L$ may be replaced by $\max\{0,-r\}$.
By shifting the summation index $L\to L-r$ it thus follows that 
$\psi_{\ell,r}(q)=\psi_{\ell,-r}(q)$.
By \eqref{psichi} this implies
$\chi_{\ell,r-\ell}(q)=\chi_{\ell,-r-\ell}(q)$ thereby
establishing the previously claimed symmetry \eqref{sym}.

The next proposition provides a fermionic representation for the second 
line in the above expression.
\begin{proposition}\label{prop2}
For $L,M$ integers and $\ell\in\{0,\dots,k\}$ such that
\begin{subequations}\label{eqp2}
\begin{equation}\label{eqp2a}
L+M+\ell\equiv 0 \pmod{2}
\end{equation}
there holds
\begin{multline}\label{eqp2b}
\sum_{\substack{i=0 \\ i\equiv L \ppmod{2}}}^k
X_{1,i+1}^{(1,k+2)}(M,\ell+1;q^{-1})X_{1,i+1}^{(1,k+2)}(L,1;q) \\
=q^{(L^2-(M+\ell)^2)/(4k)-(L-M-\ell)/2}\!
\sum_{\substack{n\in\Z^{k-1} \\ \frac{L+M+\ell}{2k}+(C^{-1}n)_1\in\Z}}\!
q^{nC^{-1}(n-Me_1-e_{\ell})}\prod_{i=1}^{k-1}\qbin{n_i+m_i}{n_i}.
\end{multline}
Here the $m_i$ follow from 
\begin{equation}\label{eqp2c}
m=C^{-1}\Bigl[(L+M)e_1+e_{\ell}-2n\Bigr].
\end{equation}
\end{subequations}
\end{proposition}

Clearly, combining Proposition~\ref{prop1} (with $n\to n_0$)
and Proposition~\ref{prop2} (with $M\to n_0$ and $L\to 2L+r$) 
results in 
\begin{equation*}
\psi_{\ell,r}^{\Bup}(q)=\psi_{\ell,r}^{\Sup}(q).
\end{equation*}

Our third and final proposition is essentially a formula from \cite{SW98}.
\begin{proposition}\label{prop3}
For $r,n_0$ integers and $\ell=\{0,\dots,k\}$ such that
\begin{equation*}
n_0+r+\ell\equiv 0\pmod{2}
\end{equation*}
there holds
\begin{multline*}
\sum_{L=0}^{\infty}\frac{q^{L(L+r)/k}}{(q;q)_L(q;q)_{L+r}}
\sum_{\substack{n\in\Z^{k-1} \\ \frac{2L+r+\ell+n_0}{2k}+(C^{-1}n)_1\in\Z}}
q^{nC^{-1}(n-n_0 e_1-e_{\ell})}\prod_{i=1}^{k-1}\qbin{n_i+m_i}{n_i} \\
=\frac{1}{(q;q)_{\infty}}
\sum_{\substack{n\in\Z^{k-1} \\ \frac{r+\ell+n_0}{2k}+(C^{-1}n)_1\in\Z}}
\frac{q^{nC^{-1}(n-n_0 e_1-e_{\ell})}}{(q;q)_{n_1}\cdots (q;q)_{n_{k-1}}},
\end{multline*}
with the $m_i$ determined by \eqref{mn}.
\end{proposition}

Applying the above to the first expression on the right of 
\eqref{BF}, i.e., to $\psi_{\ell,r}^{\Sup}(q)$, yields
\begin{equation*}
\psi_{\ell,r}^{\Sup}(q)=\psi_{\ell,r}^{\Fup}(q).
\end{equation*}

\subsection{From \eqref{formula} to \eqref{formula2}}\label{sec0}
In this preliminary section, which is not part of the proof
of Theorem~\ref{thm1}, we complete our earlier discussion
and show how the coset decomposition \eqref{cosetdec}
naturally gives rise to the character expression \eqref{formula2}.
This in particular motivates Proposition~\ref{prop1} and, 
to a lesser extend, Proposition~\ref{prop2}.

We already sketched (no actual proof of \eqref{noproof}
has been given) how \eqref{cosetdec} leads to
\eqref{formula}. 
To transform this into \eqref{formula2} we need to express
both characters in the summand on the right in terms of
one-dimensional configuration sums.

In the case of the Virasoro characters 
$\tilde{\chi}_{r,s}^{(p,p')}(q)$ this is not difficult.
Of course the simplest connection between Virasoro characters
and configuration sums follows by taking the large $L$ limit
in the latter:
\begin{equation*}
\chi_{r,s}^{(p,p')}(q)=\lim_{L\to\infty}
X^{(p,p')}_{r,s}(L,b;q)=
\sum_{j=-\infty}^{\infty} \Bigl\{
q^{j(pp'j+p'r-ps)}-q^{(pj+r)(p'j+s)}\Bigr\}.
\end{equation*}
Here $\chi_{r,s}^{(p,p')}(q)=1+O(q)$ is the normalised
character
\begin{equation}\label{norm}
\tilde{\chi}_{r,s}^{(p,p')}(q)=
q^{h_{r,s}^{(p,p')}-c^{(p,p')}/24}
\chi_{r,s}^{(p,p')}(q),
\end{equation}
with 
\begin{align*}
c^{(p,p')}&=1-\frac{6(p-p')^2}{pp'} \\
h_{r,s}^{(p,p)}&=\frac{(p'r-ps)^2-(p'-p)^2}{4pp'}
\end{align*}
the central charge and conformal weights of the
minimal model $\Mc(p,p')$.

There is however another way in which the Virasoro 
characters follow from the configuration sums, 
and by a straightforward application
of \eqref{Durfee} and 
\begin{equation*}
\qbin{n}{k}_{q^{-1}}=q^{-k(n-k)}\qbin{n}{k}_q
\end{equation*}
one finds
\begin{equation}\label{cX}
\chi_{s,2b-r}^{(p',2p'-p)}(q)=
q^{-(s-b)^2/2}
\sum_{\substack{n=0 \\ 
n\equiv s+b\ppmod{2}}}^{\infty}
\frac{q^{n^2/2}}{(q;q)_n} 
X_{r,s}^{(p,p')}(n,b;q^{-1}).
\end{equation}

The problem of expressing the parafermion character 
$\hat{\chi}_{\ell,r}(q)$ in terms of configuration sums is 
a lot more difficult, but has been
fully resolved in \cite[Cor. 4.2]{SW02}.
Using this result with $p=1$ and $p'=k+2$, we find
\begin{equation*}
\hat{b}_{i,r}(q)=(q;q)_{\infty}
q^{\Delta_i+(i^2-r^2)/(4k)}
\sum_{L=0}^{\infty}
\frac{X_{0,i+1}^{(p,p')}(2L+r,1;q)}{(q;q)_L(q;q)_{L+r}}
\end{equation*}
with $\Delta_i$ given by \eqref{Deltaell}.
Also using \cite[Eq. (2.9)]{SW02}
\begin{equation}\label{XtoX}
X_{0,s}^{(p,p')}(L,1;q)=q^{(L-s+1)/2}X_{1,s}^{(p,p')}(L,1;q)
\end{equation} 
and \eqref{bchihat} this yields
\begin{equation*}
\hat{\chi}_{i,\frac{r-i}{2}}(q)=
(q;q)_{\infty}q^{\Delta_i+(i^2-r^2)/(4k)}
\sum_{L=0}^{\infty}
\frac{q^{(2L+r-i)/2}
X_{1,i+1}^{(1,k+2)}(2L+r,1;q)}{(q;q)_L(q;q)_{L+r}}.
\end{equation*}

Substituting this as well as \eqref{cX} 
(with $p=1$, $p'=k+2$, $r=1$, $s=i+1$ and $b=\ell+1$) 
into \eqref{formula} (to this end we also need \eqref{norm})
results in \eqref{formula2} since
\begin{equation*}
h_{i+1,2\ell+1}^{(k+2,2k+3)}-\frac{1}{24}c^{(k+2,2k+3)}
+\Delta_i+\frac{i^2}{4k}
-\binom{\ell-i}{2}
=h_{\ell}+\frac{\ell^2}{4k}-\frac{c}{24}.
\end{equation*}

\subsection{Proof of Proposition~\ref{prop1}}\label{secprop1}
By \eqref{cX} and \eqref{XtoX} 
Proposition~\ref{prop1} can be simplified to
\begin{multline*}
\psi_{\ell,r}^{\Bup}(q)=(q;q)_{\infty} 
\sum_{L=0}^{\infty}\frac{1}{(q;q)_L(q;q)_{L+r}}
\\ \times
\sum_{\substack{i=0 \\ i\equiv r \ppmod{2}}}^k
q^{\binom{\ell-i}{2}}\chi_{i+1,2\ell+1}^{(k+2,2k+3)}(q)
X_{0,i+1}^{(1,k+2)}(2L+r,1;q).
\end{multline*}
Next we apply the following lemma with $p=1$ and $p'=k+2$.
\begin{lemma}\label{lem1}
For $P=p'$ and $P'=2p'-p$,
\begin{multline}\label{iden}
\sum_{\substack{i=0 \\ i\equiv r \ppmod{2}}}^{p'-2}
q^{\binom{\ell-i}{2}}\chi_{i+1,2\ell+1}^{(P,P')}(q)
X_{0,i+1}^{(p,p')}(2L+r,1;q) \\
=\frac{1}{(q;q)_{\infty}}\sum_{j,n=-\infty}^{\infty}
q^{pj(P'j+2\ell+1)/2+\binom{2n+\ell-r+P'j}{2}}
\biggl\{\qbin{2L+r}{L+n}-\qbin{2L+r}{L+n-1}\biggr\}.
\end{multline}
\end{lemma}
Hence,
\begin{multline*}
\psi_{\ell,r}^{\Bup}(q)=
\sum_{j,n=-\infty}^{\infty}
q^{j((2k+3)j+2\ell+1)/2+\binom{2n+\ell-r+(2k+3)j}{2}}
\\ \times
\sum_{L=0}^{\infty}\frac{1}{(q;q)_L(q;q)_{L+r}}
\biggl\{\qbin{2L+r}{L+n}-\qbin{2L+r}{L+n-1}\biggr\}.
\end{multline*}
This can be further transformed by yet another lemma.
\begin{lemma}\label{lem2}
For $n$ and $r$ integers, 
\begin{multline*}
\sum_{L=0}^{\infty}\frac{1}{(q;q)_L(q;q)_{L+r}}
\biggl\{\qbin{2L+r}{L+n}-\qbin{2L+r}{L+n-1}\biggr\} \\
=\frac{1}{(q;q)_{\infty}}\sum_{j=0}^{\infty} 
\biggl\{ 
\frac{q^j}{(q;q)_j(q;q)_{j-n}}-\frac{q^j}{(q;q)_j(q;q)_{j+n-r-1}} \biggr\}.
\end{multline*}
\end{lemma}
Thanks to this second lemma we are left with
\begin{multline*}
\psi_{\ell,r}^{\Bup}(q)=\frac{1}{(q;q)_{\infty}}
\sum_{j=-\infty}^{\infty} q^{j((2k+3)j+2\ell+1)/2} \\ \times
\sum_{m=0}^{\infty} \sum_{n=-\infty}^{\infty}
q^{\binom{2n+\ell-r+(2k+3)j}{2}}
\biggl\{ 
\frac{q^m}{(q;q)_m(q;q)_{m-n}}-\frac{q^m}{(q;q)_m(q;q)_{m+n-r-1}} \biggr\}.
\end{multline*}
Finally replacing $n\to m-n$ and $n\to n-m+r+1$ respectively
in the two sums over $n$ on the
right, and recalling \eqref{gradeV}, we find
\begin{equation*}
\psi_{\ell,r}^{\Bup}(q)=\frac{1}{(q;q)_{\infty}}
\sum_{j=-\infty}^{\infty} q^{j((2k+3)j+2\ell+1)/2}
\Bigl\{V_{r-\ell-(2k+3)j}(q)-V_{r+\ell+1+(2k+3)j}(q)\Bigr\}
\end{equation*}
in accordance with the left-hand side of \eqref{BF}.

Of course we still need to prove Lemmas~\ref{lem1} and \ref{lem2}.
\begin{proof}[Proof of Lemma~\ref{lem1}]
The proof of \eqref{iden} is rather simple.
In the double sum on the right we replace 
$n\to p'n+(r-i)/2$ followed by $j\to j-n$.
Hence, since $p'=P$,
\begin{align*}
\text{RHS}\eqref{iden}&=
\frac{1}{(q;q)_{\infty}}
\sum_{\substack{i=0 \\ i\equiv r \ppmod{2}}}^{2p'-2}
q^{\binom{\ell-i}{2}}
\sum_{j=-\infty}^{\infty}
q^{j(PP'j+P(2\ell+1)-P'(i+1))} \\
& \qquad \times
\sum_{n=-\infty}^{\infty}
\biggl\{q^{pn(p' n-i-1)}\biggl\{\qbin{2L+r}{(2L+r-i)/2+p'n} \\
& \qquad \qquad \qquad\qquad \qquad \qquad \qquad  -
\qbin{2L+r}{(2L+r-i-2)/2+p'n}\biggr\} \\
&=\frac{1}{(q;q)_{\infty}}
\sum_{\substack{i=0 \\ i\equiv r \ppmod{2}}}^{2p'-2}
q^{\binom{\ell-i}{2}}
X_{0,i+1}^{(p,p')}(2L+r,1;q) \\ & \qquad \qquad\qquad \qquad \times
\sum_{j=-\infty}^{\infty}
q^{j(PP'j+P(2\ell+1)-P'(i+1))}.
\end{align*}
Next we use
\begin{equation*}
\sum_{\substack{i=0 \\ i\equiv r\ppmod{2}}}^{2p'-2} f_{i+1}=
\sum_{\substack{i=0 \\ i\equiv r\ppmod{2}}}^{p'-2} (f_{i+1}+f_{2p'-i-1})
\end{equation*}
provided that $f_{p'}=0$.
Since
\begin{equation}\label{Xnul}
X_{0,p'}^{(p,p')}(L,1;q)=0
\end{equation}
this may be applied to the sum over $i$. Also using
\begin{equation*}
X_{0,i+1}^{(p,p')}(L,1;q)=-q^{p(p'-i-1)}X_{0,2p'-i-1}^{(p,p')}(L,1;q)
\end{equation*}
(this in fact implies \eqref{Xnul}) yields
\begin{multline*}
\text{RHS}\eqref{iden}=
\frac{1}{(q;q)_{\infty}}
\sum_{\substack{i=0 \\ i\equiv r \ppmod{2}}}^{p'-2}
q^{\binom{\ell-i}{2}}
X_{0,i+1}^{(p,p')}(2L+r,1;q) \\ \times
\sum_{j=-\infty}^{\infty}
\Bigl\{ q^{j(PP'j+P(2\ell+1)-P'(i+1))}
-q^{(P(j-1)+i+1)(P'(j-1)+2\ell+1)}\Bigr\}.
\end{multline*}
Replacing $j\to j+1$ in the second term in the sum over $j$ completes the proof.
\end{proof}

\begin{proof}[Proof of Lemma~\ref{lem2}]
We will show that
\begin{equation}\label{ab}
(a-b)\sum_{j=0}^{\infty} 
\frac{(ab;q)_{2j}\,q^j}{(q,aq,bq,ab;q)_j}
=\sum_{j=0}^{\infty} q^j\biggl\{
\frac{a}{(q,aq;q)_j(bq;q)_{\infty}}-
\frac{b}{(q,bq;q)_j(aq;q)_{\infty}}\biggr\}.
\end{equation}
Letting $a\to aq^n$, $b\to bq^{r-n+1}$ this gives
\begin{multline*}
\sum_{j=0}^{\infty} 
\frac{(abq;q)_{2j+r} (aq^{j+n}-bq^{j+r-n+1})}
{(q;q)_j(abq;q)_{j+r}(aq;q)_{j+n}(bq;q)_{j+r-n+1}} \\
=\sum_{j=0}^{\infty} \biggl\{
\frac{aq^{j+n}}{(q;q)_j(aq;q)_{j+n}(bq;q)_{\infty}}-
\frac{bq^{j+r-n+1}}{(q;q)_j(bq;q)_{j+r-n+1}(aq;q)_{\infty}}\biggr\}.
\end{multline*}
Shifting $j\to j-n$ and $j\to j-r+n-1$ in the two terms on the right
and putting $a=b=1$ yields the claim of the lemma since
\begin{equation*}
\frac{(q^{j+n}-q^{j+r-n+1})(q;q)_{2j+r}}{(q;q)_{j+n}(q;q)_{j+r-n+1}}
=\qbin{2j+r}{j+n}-\qbin{2j+r}{j+n-1}.
\end{equation*}

\medskip

Equation \eqref{ab} is similar to \cite[Eq. (4.3)]{SW02}
and \cite[Thm 1.5]{W03} and the proof proceeds accordingly.
First we multiply both sides of \eqref{ab} by $(aq,bq;q)_{\infty}$
and use $(a;q)_{\infty}/(a;q)_n=(aq^n;q)_{\infty}$ to obtain
\begin{equation*}
(a-b)\sum_{j=0}^{\infty} \frac{q^j (abq^j;q)_j(aq^{j+1},bq^{j+1};q)_{\infty}}
{(q;q)_j} \
=\sum_{j=0}^{\infty} q^j\biggl\{
\frac{a(aq^{j+1};q)_{\infty}}{(q;q)_j}-
\frac{b(bq^{j+1};q)_{\infty}}{(q;q)_j}\biggr\}.
\end{equation*}
Next we expand each of the $q$-shifted factorials depending on $a$ and/or $b$
by \eqref{eulera} or the $q$-binomial theorem \cite[Eq. (3.3.6)]{Andr}
\begin{equation*}
\sum_{n=0}^{\infty} (-1)^n a^n q^{\binom{n}{2}}\qbin{M}{n}=(a;q)_M.
\end{equation*}
This results in 
\begin{multline*}
\sum_{j,k,l,n=0}^{\infty} 
\frac{(-1)^{k+l+n}a^{k+n} b^{l+n}
q^{\binom{k+1}{2}+\binom{l+1}{2}+\binom{n}{2}+j(k+l+n+1)}}
{(q;q)_n(q;q)_{j-n}(q;q)_k(q;q)_l}\\
=\sum_{j,k=0}^{\infty} 
\frac{a^{k+1}-b^{k+1}}{a-b}
\frac{(-1)^kq^{\binom{k+1}{2}+j(k+1)}}{(q;q)_j(q;q)_k}.
\end{multline*}
After the shift $j\to j+n$ on the left, both sums over $j$ can be carried
out by \eqref{euler}, leading to
\begin{multline*}
\sum_{k,l,n=0}^{\infty} 
\frac{(-1)^{k+l+n}a^{k+n} b^{l+n}
q^{\binom{k+1}{2}+\binom{l+1}{2}+\binom{n}{2}+n(k+l+n+1)}(q;q)_{k+l+n}}
{(q;q)_n(q;q)_k(q;q)_l}\\
=\sum_{k=0}^{\infty} 
\frac{a^{k+1}-b^{k+1}}{a-b}(-1)^kq^{\binom{k+1}{2}}.
\end{multline*}
Equating coefficients of $a^k b^l$ and
performing some standard manipulations yields
\begin{equation*}
{_2\phi_1}(q^{-k},q^{-l};q^{-k-l};q,1)
=q^{kl}\frac{(q;q)_k(q;q)_l}{(q;q)_{k+l}}.
\end{equation*}
Here 
\begin{multline*}
{_{r+1}\phi_r}\biggl[\genfrac{}{}{0pt}{}{a_1,\cdots,
a_{r+1}}{b_1,\cdots,b_r};q,z\biggr]  \\
={_{r+1}\phi_r}(a_1,\cdots,a_{r+1};b_1,\cdots,b_r;q,z)
=\sum_{n=0}^{\infty}
\frac{(a_1,\cdots,a_{r+1};q)_n}{(b_1\cdots,b_r,q;q)_n}z^n
\end{multline*}
is a basic hypergeometric series, see \cite{GR}.
Summing the ${_2\phi_1}$ series using the $q$-Chu--Vandermonde sum
\cite[Eq. (II.7)]{GR}
\begin{equation*}
{_2\phi_1}(a,q^{-n};c;q,cq^n/a)=\frac{(c/a;q)_n}{(c;q)_n}
\end{equation*}
completes the proof.
\end{proof}

It is interesting to note that \eqref{ab} admits the following
bounded analogue
\begin{multline*}
(a-b)\sum_{j=0}^M q^j \qbin{M}{j}
\frac{(ab;q)_{2j}}{(aq,bq,ab;q)_j} \\
=\sum_{j=0}^M q^j\qbin{M}{j}\biggl\{
\frac{a}{(aq;q)_j(bq;q)_{M-j}}-
\frac{b}{(bq;q)_j(aq;q)_{M-j}}\biggr\}.
\end{multline*}
We leave its proof (which is surprisingly difficult) to the reader.

\subsection{Proof of Proposition~\ref{prop2}}\label{secprop2}

In \cite{W96a,W96b} a combinatorial technique was developed
to obtain fermionic representations for the
one-dimensional configuration sums
\begin{equation*}
X_{r,s}^{(p,p+1)}(L,b;q).
\end{equation*}
Combining this technique with the work of Berkovich and Paule 
\cite{BP02} on generalizations of 
the Andrews--Gordon identities results in Proposition~\ref{prop2}.
More precisely, we will first show --- following the method 
of \cite{BP02} --- that (after a simple rewriting) the left-hand 
side of \eqref{eqp2b} may be interpreted as the
generating function of a particular set of lattice paths.
We will then use the method of \cite{W96a,W96b} to show that this
generating function permits a fermionic form in accordance with
the right-hand side of \eqref{eqp2b}.
The details of the proof marginally differ according
to whether $\ell=0$ or $\ell\in\{1,\dots,k\}$. 
We will first treat the $\ell=0$ case in full detail 
and then point out the relevant differences with $\ell\in\{1,\dots,k\}$.

Our initial step is to take \eqref{eqp2b} with $\ell=0$ and use 
\cite[Eq. (2.3)]{SW02}
\begin{equation}\label{Xinv}
X_{r,s}^{(p,p')}(L,b;q)=
q^{(L^2-(b-s)^2)/4} X_{b-r,s}^{(p'-p,p')}(L,b;q^{-1})
\end{equation}
followed by \eqref{XtoX} to transform the left-hand side. 
On the right-hand side we use \eqref{eqp2c} to
eliminate $n$ in the exponent of $q$ in 
favour of $m$. As a result we obtain the $\ell=0$ instance of
the more general identity
\begin{subequations}\label{Pee}
\begin{multline}\label{P1}
\sum_{\substack{i=0 \\ i\equiv L \ppmod{2}}}^k
q^{\ell(\ell-2i+1)/4}
X_{\ell+1,i+1}^{(k+1,k+2)}(M,\ell+1;q)X_{1,i+1}^{(k+1,k+2)}(L,1;q^{-1}) \\
=\sum_{\substack{n\in\Z^{k-1} \\ \frac{L+M-\ell}{2k}+(C^{-1}n)_1\in\Z}}
q^{\frac{1}{4}mCm-\frac{1}{2}Lm_1}\prod_{i=1}^{k-1}\qbin{n_i+m_i}{n_i},
\end{multline}
where
\begin{equation}\label{P2}
m=C^{-1}\Bigl[(L+M)e_1+e_{k-\ell}-2n\Bigr]
\end{equation}
and
\begin{equation}\label{P3}
L+M+\ell\equiv 0 \pmod{2}
\end{equation}
and 
\begin{equation}\label{P4}
\ell\in\{0,\dots,k-1\}.
\end{equation}
\end{subequations}

\begin{figure}[tb]
\begin{center}
\begin{pspicture}(0,0)(12,4)
\psline{->}(2.5,0.5)(2.5,3.5)
\psline{->}(0.5,0.5)(11.5,0.5)
\psline{-}(0.5,0.5)(0.5,0.6)
\psline{-}(1.0,0.5)(1.0,0.6)
\psline{-}(1.5,0.5)(1.5,0.6)
\psline{-}(2.0,0.5)(2.0,0.6)
\psline{-}(2.5,0.5)(2.5,0.6)
\psline{-}(3.0,0.5)(3.0,0.6)
\psline{-}(3.5,0.5)(3.5,0.6)
\psline{-}(4.0,0.5)(4.0,0.6)
\psline{-}(4.5,0.5)(4.5,0.6)
\psline{-}(5.0,0.5)(5.0,0.6)
\psline{-}(5.5,0.5)(5.5,0.6)
\psline{-}(6.0,0.5)(6.0,0.6)
\psline{-}(6.5,0.5)(6.5,0.6)
\psline{-}(7.0,0.5)(7.0,0.6)
\psline{-}(7.5,0.5)(7.5,0.6)
\psline{-}(8.0,0.5)(8.0,0.6)
\psline{-}(8.5,0.5)(8.5,0.6)
\psline{-}(9.0,0.5)(9.0,0.6)
\psline{-}(9.5,0.5)(9.5,0.6)
\psline{-}(10.0,0.5)(10.0,0.6)
\psline{-}(10.5,0.5)(10.5,0.6)
\psline{-}(11.0,0.5)(11.0,0.6)
\rput(2.5,0.25){{\small $0$}}
\rput(3.0,0.25){{\small $1$}}
\rput(5.0,0.25){{\small $5$}}
\rput(7.5,0.25){{\small $10$}}
\psline{-}(2.5,1.0)(2.6,1.0)
\psline{-}(2.5,1.5)(2.6,1.5)
\psline{-}(2.5,2.0)(2.6,2.0)
\psline{-}(2.5,2.5)(2.6,2.5)
\psline{-}(2.5,3.0)(2.6,3.0)
\rput(2.25,1.0){{\small $1$}}
\rput(2.25,2.0){{\small$\ell$}}
\rput(2.25,3.0){{\small $5$}}
\psline{-}(1.0,0.5)(1.5,1.0)
\psline{-}(1.5,1.0)(2.0,0.5)
\psline{-}(2.0,0.5)(3.0,1.5)
\psline{-}(3.0,1.5)(3.5,1.0)
\psline{-}(3.5,1.0)(5.0,2.5)
\psline{-}(5.0,2.5)(7.0,0.5)
\psline{-}(7.0,0.5)(9.0,2.5)
\psline{-}(9.0,2.5)(10.0,1.5)
\psline{-}(10.0,1.5)(11.0,2.5)
\rput(10.5,0.25){{\small $M$}} 
\end{pspicture}
\end{center}
\caption{Example of path $P$ with $k\geq 4$, $L=3$, $M=16$ and
$\ell=3$.}\label{figure1}
\end{figure}
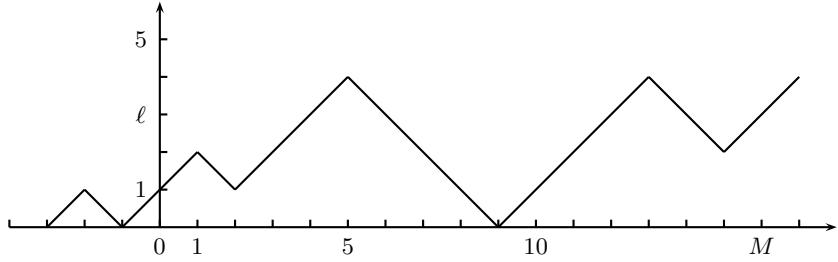

What we will set out to do is to compute the generating function
of lattice paths contained in a strip of height $k$ starting in
$(-L,0)$ and terminating with a step from $(M,\ell)$ to $(M+1,\ell+1)$.
More precisely, starting from $(-L,0)$ we carry out a sequence of 
north-east (up or $+$) and south-east (down or $-$) steps 
(an up step is from $(x,y)$ to $(x+1,y+1)$ and a down step is from
$(x,y)$ to $(x+1,y-1)$, with $x,y$ integers)
such that the $y$ coordinate of each
vertex along the path is always nonnegative and never exceeds $k$,
and such that the last step is in the north-east direction and
terminates in $(M+1,\ell+1)$.
An example of such a lattice path is shown in Figure~\ref{figure1}.
Obviously there are no admissible paths unless
\eqref{P3} and \eqref{P4} are satisfied.
The \textit{shape} of a path is given by its
sequence of consecutive steps, ignoring the actual
positions of the path's starting and ending vertices.
The shape of the path of Figure~\ref{figure1} is 
$$(+,-,+,+,-,+,+,+,-,-,-,-,+,+,+,+,-,-,+,+).$$

\begin{figure}[tb]
\begin{center}
\begin{pspicture}(0,0)(12,4)
\psline{-}(1.5,0.5)(3.5,0.5)
\psline{-}(7.0,0.5)(9.0,0.5)
\psline{-}(1.5,0.5)(1.5,0.6)
\psline{-}(2.0,0.5)(2.0,0.6)
\psline{-}(2.5,0.5)(2.5,0.6)
\psline{-}(3.0,0.5)(3.0,0.6)
\psline{-}(3.5,0.5)(3.5,0.6)
\psline{-}(7.0,0.5)(7.0,0.6)
\psline{-}(7.5,0.5)(7.5,0.6)
\psline{-}(8.0,0.5)(8.0,0.6)
\psline{-}(8.5,0.5)(8.5,0.6)
\psline{-}(9.0,0.5)(9.0,0.6)
\psline{*-*}(2.0,1.5)(2.5,2.0)
\psline{*-*}(2.5,2.0)(3.0,2.5)
\psline[linestyle=dashed]{-}(2.5,0.5)(2.5,2.0)
\psline{*-*}(7.5,2.5)(8.0,2.0)
\psline{*-*}(8.0,2.0)(8.5,1.5)
\psline[linestyle=dashed]{-}(8.0,0.5)(8.0,2.0)
\rput(2.5,0.25){{\small $x$}}
\rput(8.0,0.25){{\small $x$}}
\end{pspicture}
\end{center}
\caption{Vertices of weight $x$.}\label{figure2}
\end{figure}
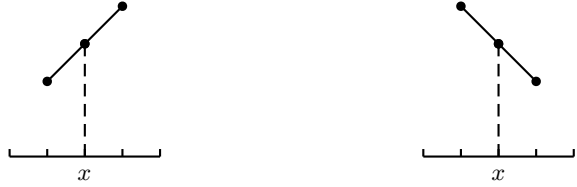

To a path $P$ we assign a weight (or statistic) $W(P)$ given by
the sum of the $x$-coordinates of vertices of the type shown in 
Figure~\ref{figure2}. For example, the weight $P$ of the path of 
Figure~\ref{figure1}
is given by 
$$0+3+4+6+7+8+10+11+12+14+16.$$
The aim is to compute the generating function
\begin{equation*}
G_{\ell}(L,M;q)=\sum_{P} q^{W(P)/2},
\end{equation*}
where the sum is over all admissible paths $P$.
The rationale behind the half in the exponent on the right 
is the fact that $W(P)\equiv \binom{\ell+1}{2}-\ell L\pmod{2}$
so that, up to a possible overall factor $q^{1/2}$, 
$G_{\ell}(L,M;q)$ is a Laurent polynomial in $q$.

Our first method of computing $G_{\ell}(L,M;q)$ closely follows 
\cite{BP02}, and, as we shall see shortly, gives the left-hand 
side of \eqref{P1}.
First let us introduce the analogous problem of computing lattice paths 
in the strip with the same bounds on the $y$ coordinates but with initial
vertex $(0,i)$ and final two vertices $(L,\ell)$, $(L+1,\ell+1)$.
Defining the same statistic as before we denote the generating
function of such paths by $G_{i,\ell}(L;q)$.

After $L$ steps a path $P$ from $(-L,0)$ to $(M,\ell)$, $(M+1,\ell+1)$
crosses the nonnegative $y$-axis at some vertex $(0,i)$,
where $0\leq i\leq k$ and $i\equiv L\pmod{2}$.
Because this vertex has $x$-coordinate equal to zero, the `local shape' 
(i.e., $(+,+)$, $(+,-)$, $(-,+)$ or $(-,-)$) of the paths when 
crossing the $y$-axis will not affect the weight $W(P)$. 
Consequently, one may view $P$ as the concatenation of two lattice
paths $P_{+}$ and $P_{-}$,
the former starting at $(0,i)$, pointing in the positive
$x$-direction and terminating in $(M,\ell)$, $(M+1,\ell+1)$ 
and the latter starting at $(0,i)$, pointing in the negative $x$-direction,
and terminating in $(-L,0)$, $(-L-1,1)$.
(The addition of the final up-step from $(-L,0)$ to $(-L-1,1)$
is permitted since it will not change the weight of $P_{-}$. Indeed the 
final two steps of $P_{-}$ will now be $(-,+)$).
The path $P_{-}$ may be reflected in the $y$-axis so that it
becomes a path pointing in the positive $x$-direction,
starting at $(0,i)$ and terminating in 
$(L,0)$ to $(L+1,1)$. Since the reflection negates the weight
of $P_{-}$ we need to replace $q$ by $q^{-1}$ in the corresponding
generating function of $P_{-}$ type paths.
At the level of the generating function the above
decomposition thus becomes
\begin{equation*}
G_{\ell}(L,M;q)=\sum_{\substack{i=0 \\ i\equiv L\ppmod{2}}}^k 
G_{i,\ell}(M;q)G_{i,0}(L;q^{-1}).
\end{equation*}
Since $G_{i,\ell}(0;q)=\delta_{i,\ell}$ this includes the obvious
relations
\begin{equation*}
G_{\ell}(0,M;q)=G_{0,\ell}(M;q) \quad \text{and} \quad
G_{\ell}(L,0;q)=G_{\ell,0}(L;q^{-1}).
\end{equation*}

The $G_{i,\ell}(L;q)$ may be computed from a simple recursion, 
and quoting the result of \cite{ABF84} we have
\begin{equation}\label{ABF1}
G_{i,\ell}(L;q)=q^{(i-\ell)(i-\ell-1)/4}
X^{(k+1,k+2)}_{\ell+1,i+1}(L,\ell+1;q).
\end{equation} 
The generating function $G_{\ell}(L,M;q)$ is thus found to be
\begin{equation*}
G_{\ell}(L,M;q)=\sum_{\substack{i=0 \\ i\equiv L\ppmod{2}}}^k 
q^{\ell(\ell-2i+1)/4}
X^{(k+1,k+2)}_{\ell+1,i+1}(M,\ell+1;q)
X^{(k+1,k+2)}_{1,i+1}(L,1;q^{-1})
\end{equation*}
in accordance with the left-hand side of \eqref{P1}.

The second method of computing $G_{\ell}(L,M;q)$
closely follows \cite{W96a,W96b} where it was used to
find fermionic representations for 
the generating function $G_{i,\ell}(L;q)$.
The method amounts to interpreting an admissible path
as a collection of charged particles with charges taken from the
set $\{1,\dots,k\}$
(the charge being the effective height of a particle).
Hence each admissible path $P$ may be assigned
a sequence of integers $(n_1,\dots,n_k)$ where $n_j$ is the
number of particles of charge $j$ in $P$.
For example, the decomposition into particles of the path 
in Figure~\ref{figure3}
is indicated by the dotted lines, and this particular path has
$n_1=2$, $n_2=5$ and $n_4=1$ and all other $n_j=0$.
Because a path has fixed initial and final vertices, the $n_j$ are
subject to the constraint
\begin{equation*}
2\sum_{j=1}^k j n_j=L-i-\ell.
\end{equation*}

\begin{figure}[tb]
\begin{center}
\begin{pspicture}(0,0)(10.75,4)
\psline{-}(0.5,0.5)(0.5,1.5)
\psline{-}(0.5,1.5)(0.75,1.0)
\psline{-}(0.75,1.0)(1.25,2.0)
\psline{-}(1.25,2.0)(1.75,1.0)
\psline{-}(1.75,1.0)(2.0,0.5)
\psline{-}(2.0,0.5)(2.75,2.0)
\psline{-}(2.75,2.0)(3.0,1.5)
\psline{-}(3.0,1.5)(3.25,2.0)
\psline{-}(3.25,2.0)(3.75,1.0)
\psline{-}(3.75,1.0)(4.5,2.5)
\psline{-}(4.5,2.5)(5.5,0.5)
\psline{-}(5.5,0.5)(6.0,1.5)
\psline{-}(6.0,1.5)(6.5,0.5)
\psline{-}(6.5,0.5)(6.75,1.0)
\psline{-}(6.75,1.0)(7.25,2.0)
\psline{-}(7.25,2.0)(7.75,1.0)
\psline{-}(7.75,1.0)(8.25,2.0)
\psline{-}(8.25,2.0)(8.75,3.0)
\psline{-}(8.75,3.0)(9.0,2.5)
\psline{-}(9.0,2.5)(9.25,3.0)
\psline{-}(9.25,3.0)(9.75,2.0)
\psline{-}(9.75,2.0)(10.25,3.0)
\psline{-}(10.25,3.0)(10.25,0.5)
\psline[linestyle=dashed]{-}(0.5,0.5)(10.25,0.5)
\psline[linestyle=dashed]{-}(0.75,1.0)(1.75,1.0)
\psline[linestyle=dashed]{-}(2.25,1.0)(3.75,1.0)
\psline[linestyle=dashed]{-}(3.0,1.5)(3.5,1.5)
\psline[linestyle=dashed]{-}(6.75,1.0)(7.75,1.0)
\psline[linestyle=dashed]{-}(8.25,2.0)(9.75,2.0)
\psline[linestyle=dashed]{-}(9.0,2.5)(9.5,2.5)
\rput(1.25,1.4){{$2$}}
\rput(2.75,1.4){{$2$}}
\rput(6,0.9){{$2$}}
\rput(7.25,1.4){{$2$}}
\rput(8.75,2.4){{$2$}}
\rput(3.25,1.725){{$1$}}
\rput(9.25,2.725){{$1$}}
\rput(4.5,1.4){{$4$}}
\end{pspicture}
\end{center}
\caption{Example of path $P$ with particle content 
$(n_1,n_2,n_3,n_4,\dots)=(2,5,0,1,0,0,\dots)$.}
\label{figure3}
\end{figure}

The important point of the combinatorial method is that the 
generating function of paths with fixed sequence $(n_1,\dots,n_k)$,
i.e., with fixed particle content may easily be computed, 
and as a special case we quote \cite[Prop. 3]{W96b}
\begin{equation}\label{G0}
G_{0,\ell}(L;n_1,\dots,n_k;q)
=q^{\frac{1}{4}mCm}\prod_{j=1}^{k-1}\qbin{n_j+m_j}{n_j},
\end{equation}
where the $m_j$ follow from
\begin{equation}\label{MN}
m=C^{-1}\Bigl[Le_1+e_{k-\ell}-2n\Bigr]
\end{equation}
and where the $n_j$ are subject to
\begin{equation}\label{comp}
2\sum_{j=1}^k j n_j=L-\ell.
\end{equation}

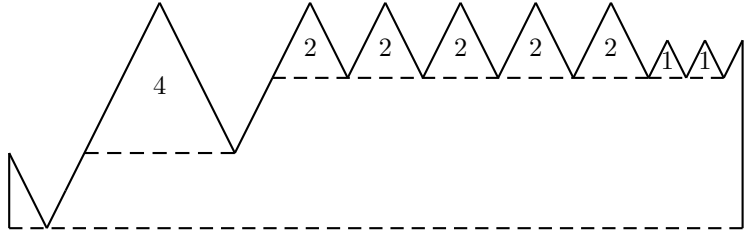
\begin{figure}[tb]
\begin{center}
\begin{pspicture}(0,0)(10.75,4)
\psline{-}(0.5,0.5)(0.5,1.5)
\psline{-}(0.5,1.5)(1.0,0.5)
\psline{-}(1.0,0.5)(1.25,1.0)
\psline{-}(1.25,1.0)(2.5,3.5)
\psline{-}(2.5,3.5)(3.5,1.5)
\psline{-}(3.5,1.5)(4.0,2.5)
\psline{-}(4.0,2.5)(4.5,3.5)
\psline{-}(4.5,3.5)(5.0,2.5)
\psline{-}(5.0,2.5)(5.5,3.5)
\psline{-}(5.5,3.5)(6.0,2.5)
\psline{-}(6.0,2.5)(6.5,3.5)
\psline{-}(6.5,3.5)(7.0,2.5)
\psline{-}(7.0,2.5)(7.5,3.5)
\psline{-}(7.5,3.5)(8.0,2.5)
\psline{-}(8.0,2.5)(8.5,3.5)
\psline{-}(8.5,3.5)(9.0,2.5)
\psline{-}(9.0,2.5)(9.25,3.0)
\psline{-}(9.25,3.0)(9.5,2.5)
\psline{-}(9.5,2.5)(9.75,3.0)
\psline{-}(9.75,3.0)(10.0,2.5)
\psline{-}(10.0,2.5)(10.25,3.0)
\psline{-}(10.25,3.0)(10.25,0.5)
\psline[linestyle=dashed]{-}(0.5,0.5)(10.25,0.5)
\psline[linestyle=dashed]{-}(1.5,1.5)(3.5,1.5)
\psline[linestyle=dashed]{-}(4.0,2.5)(9.0,2.5)
\psline[linestyle=dashed]{-}(9.0,2.5)(10.0,2.5)
\rput(4.5,2.9){{$2$}}
\rput(5.5,2.9){{$2$}}
\rput(6.5,2.9){{$2$}}
\rput(7.5,2.9){{$2$}}
\rput(8.5,2.9){{$2$}}
\rput(9.25,2.725){{$1$}}
\rput(9.75,2.725){{$1$}}
\rput(2.5,2.4){{$4$}}
\end{pspicture}
\end{center}
\caption{Minimal path corresponding to the path of Figure~\ref{figure3}
with $k=6$.
All particles are moved to the right as much as possible and arranged in
decreasing order with respect to charge, such that the maximal
height of the path does not exceed $k$ and such that
the final step remains upwards}\label{figure4}
\end{figure}

The term $\frac{1}{4}mCm$ in the above summand is the weight of the
`minimal path of content $(n_1,\dots,n_k)$' corresponding to
the special type of path shown in Figure~\ref{figure4}. All other paths with
the same content have a weight of the form $\frac{1}{4}mCm+N$
where $N$ is a positive integer, and together they generate the
product of $q$-binomials in \eqref{G0}.
The value of $N$ for a non-minimal path $P$
is given by the number of `elementary' moves
required to obtain $P$ from its corresponding minimal path $P_{\min}$,
and, importantly, \textit{only} depends on the shape of a path.
We refer the reader to \cite{W96a,W96b} for the precise details.

Let us now return to the problem of computing the generating function
$G_{\ell}(L,M;q)$. Since we can again make a particle decomposition
of admissible paths, we may first try to compute
$G_{\ell}(L,M;n_1,\dots,n_k;q)$.
This follows from 
$G_{0,\ell}(L;n_1,\dots,n_k;q)$
by carrying out the following two transformations, the first acting on the
generating function and the second acting on the actual paths.
\begin{enumerate}
\item
Replace $L$ by $L+M$ in $G_{0,\ell}(L;n_1,\dots,n_k;q)$, so that
we are counting paths from $(0,0)$ to $(L+M,\ell)$, $(L+M+1,\ell+1)$.
\item
Translate each path counted by $G_{0,\ell}(L+M;n_1,\dots,n_k;q)$
exactly $L$ units to the left.
\end{enumerate}
The first transformation simply means that 
\eqref{MN} and \eqref{comp} need to be replaced by \eqref{P2} and
\begin{equation}\label{comp2}
2\sum_{j=1}^k j n_j=L+M-\ell.
\end{equation}
The second transformation has a more subtle effect.
The weight of a path is computed by summing up the $x$ coordinates of 
$(+,+)$ and $(-,-)$ sequences of steps, and
it is non-trivial to relate the weight of a left-translated path to 
its weight before translation.
However, the crucial observation is that the weights of
all paths with the same content $(n_1,\dots,n_k)$ are rescaled identically. 
Specifically, if the weight of the minimal path changes as
\begin{equation*}
K\to K+\Delta K
\end{equation*}
as a result of the translation,
then a non-minimal path of the same particle content 
changes as
\begin{equation*}
K+N\to K+N+\Delta K.
\end{equation*}
This follows from the previously-mentioned facts that (i) the weight of 
a non-minimal path $P$ is obtained from its minimal path 
$P_{\min}$
by counting the number, $N$, of elementary moves required to obtain 
$P$ from $P_{\min}$, and (ii) $N$ only depends on the shape of
$P$, and is thus invariant under translations.

So all we have to do is (re)calculate the weight of the minimal path
of content $(n_1,\dots,n_k)$ after its translation to the left.
This is given by $E$ with 
\begin{align*}
E&=\sum_{j=1}^kj(j-1)n_j^2+
2\sum_{j=1}^k\sum_{l=j+1}^k(j-1)ln_jn_l
+\ell\sum_{j=1}^k(j-1)n_j \\ &\quad +
\sum_{j=k-\ell+1}^k(\ell-k+j)n_j+
\frac{1}{4}\ell(\ell+1)-
L\biggl[\sum_{j=1}^k(j-1)n_j-\frac{1}{2}\ell\biggr].
\end{align*}
When $L=0$ (no left-translation)
this is the weight of the minimal paths computed in 
\cite[Eq. (3.13)]{W96b}.
Eliminating $n_k$ using the restriction \eqref{comp2},
and then writing the result in terms of the $m_j$ instead of $n_j$
using \eqref{P2}, yields
\begin{equation*}
E=\frac{1}{4}mCm-\frac{1}{2}Lm_1.
\end{equation*}
We therefore conclude that
\begin{equation}\label{Gee}
G_{\ell}(L,M;n_1,\dots,n_k;q)
=q^{\frac{1}{4}mCm-\frac{1}{2}Lm_1}\prod_{j=1}^{k-1}\qbin{n_j+m_j}{n_j},
\end{equation}
where the $m_j$ follow from \eqref{P2} and where the $n_j$ must satisfy
\eqref{comp2}.
To obtain the full generating function all we need to do is sum over all
admissible sequences $(n_1,\dots,n_k)$.
Now \eqref{comp2} may be rewritten as
\begin{equation*}
\frac{L+M-\ell}{2k}+(C^{-1}n)_{1}=n_1+n_2+\cdots+n_k
\end{equation*}
with $n=(n_1,\dots,n_{k-1})$.
Since the right-hand side of \eqref{Gee} has no explicit dependence
on $n_k$, summing over admissible sequences simply corresponds to summing
over $n\in\Z^{k-1}$ such that
\begin{equation*}
\frac{L+M-\ell}{2k}+(C^{-1}n)_{1}\in\Z.
\end{equation*}
In conclusion, $G_{\ell}(L,M;q)$ is given by the 
right-hand side of \eqref{P1} completing the proof of \eqref{Pee}.

\medskip

Finally we need to return to \eqref{eqp2b} with $\ell\in\{1,\dots,k\}$.
Using \eqref{XtoX}, \eqref{Xinv} and
\begin{equation*}
X_{r,s}^{(p,p')}(L,b;q)=X_{p-r,p'-s}^{(p,p')}(L,p'-b;q)
\end{equation*}
on the left-hand side and eliminating $n$ in the exponent of $q$
on the right-hand side we find the following variation on \eqref{P1}:
\begin{multline}\label{Q1}
\sum_{\substack{i=0 \\ i\equiv L \ppmod{2}}}^k
q^{(\ell-1)(\ell-2i)/4}
X_{k-\ell+1,k-i+1}^{(k+1,k+2)}(M,k-\ell+1;q)
X_{k,k-i+1}^{(k+1,k+2)}(L,k+1;q^{-1}) \\
=\sum_{\substack{n\in\Z^{k-1} \\ \frac{L+M+\ell}{2k}+(C^{-1}n)_1\in\Z}}
q^{\frac{1}{4}mCm-\frac{1}{2}Lm_1}\prod_{i=1}^{k-1}\qbin{n_i+m_i}{n_i},
\end{multline}
subject to \eqref{eqp2a} and \eqref{eqp2c}.

This time both sides of \eqref{Q1} may be
interpreted as the generating function $H_{\ell}(L,M;q)$
of lattice paths confined to the strip with weight function as before,
but with initial vertex $(-L,k)$ and final two vertices 
$(M,k-\ell)$, $(M+1,k-\ell+1)$. 
Obviously there are no admissible paths unless
\eqref{eqp2a} and \eqref{eqp2c} are satisfied.

Previously we defined the generating function $G_{i,\ell}(L;q)$
of paths in the strip with initial
vertex $(0,i)$ and final two vertices $(L,\ell)$, $(L+1,\ell+1)$.
Let us now also define $H_{i,\ell}(L;q)$ as the generating function
of paths in the strip with initial vertex $(0,i)$ and final 
two vertices $(L,\ell)$, $(L+1,\ell-1)$.

Again a path $P$ counted by $H_{\ell}(L,M;q)$ may be seen as
the concatenation of a left-pointing path $P_{-}$ from $(0,k-i)$ to
$(-L,k)$ and a right-pointing path $P_{+}$ from $(0,k-i)$ to
$(M,k-\ell)$, $(M+1,k-\ell+1)$, where $k\in\{0,\dots,\ell\}$ and
$i\equiv L\pmod{2}$.
This time we may add a down step from $(-L,k)$ 
to $(-L-1,k-1)$ to $P_{-}$ without changing its weight
(so that $P_{-}$ terminates with a $(+,-)$ pair of steps).
Reflecting $P_{-}$ in the $y$-axis we thus find
\begin{equation*}
H_{\ell}(L,M;q)=\sum_{\substack{i=0 \\ i\equiv L\pmod{2}}}^k 
G_{k-i,k-\ell}(M;q)H_{k-i,k}(L;q^{-1}).
\end{equation*}
This includes the trivial relations
\begin{equation*}
H_{\ell}(0,M;q)=G_{k,k-\ell}(M;q) \quad \text{and} \quad
H_{\ell}(L,0;q)=H_{k-\ell,k}(L;q^{-1}).
\end{equation*}
Once more using \eqref{ABF1} as well as \cite{ABF84}
\begin{equation*}
H_{i,\ell}(L;q)=
q^{(i-\ell)(i-\ell+1)/4}X^{(k+1,k+2)}_{\ell,i+1}(L,\ell+1;q),
\end{equation*} 
yields
\begin{multline*}
H_{\ell}(L,M;q) \\
=\sum_{\substack{i=0 \\ i\equiv L\ppmod{2}}}^k 
q^{(\ell-1)(\ell-2i)/4}
X^{(k+1,k+2)}_{k-\ell+1,k-i+1}(M,k-\ell+1;q)
X^{(k+1,k+2)}_{k,k-i+1}(L,k+1;q^{-1}),
\end{multline*}
in accordance with the left-hand side of \eqref{Q1}.

To obtain a fermionic evaluation of $H_{\ell}(L,M;q)$ we need
\cite[Prop. 3]{W96b}
\begin{equation*}
G_{k,k-\ell}(L;n_1,\dots,n_k;q)
=q^{\frac{1}{4}mCm}\prod_{j=1}^{k-1}\qbin{n_j+m_j}{n_j},
\end{equation*}
where the $m_j$ follow from
\begin{equation}\label{mn3}
m=C^{-1}\Bigl[Le_1+e_{\ell}-2n\Bigr]
\end{equation}
and the $n_j$ are constraint by
\begin{equation}\label{comp3}
2\sum_{j=1}^k j n_j=L+\ell-2k.
\end{equation}
Again the $\frac{1}{4}mCm$ in the above summand is the weight of the
`minimal path of content $(n_1,\dots,n_k)$'.

To obtain $H_{\ell}(L,M;q)$ we again replace $L\to L+M$ in
the expression for $G_{k,k-\ell}(L;n_1,\dots,n_k;q)$ and then translate
all paths by $L$ units to the left.
As before, the weight $E$ of the minimal path of content
$(n_1,\dots,n_k)$ needs to be recalculated, yielding
\begin{align*}
E&=\sum_{j=1}^kj(j-1)n_j^2+
2\sum_{j=1}^k\sum_{l=j+1}^k(j-1)ln_l
+(2k-\ell)\sum_{j=1}^k(j-1)n_j \\ &\quad +
\sum_{j=\ell+1}^k(j-\ell)n_j+
\frac{1}{4}k(k-1)+\frac{1}{4}(k-\ell)(3k-\ell+1) \\
&\quad - L\biggl[\sum_{j=1}^k(j-1)n_j-\frac{1}{2}(\ell+1)\biggr].
\end{align*}
By \eqref{mn3} and \eqref{comp3} with $L\to L+M$ this yields
\begin{equation*}
E=\frac{1}{4}mCm-\frac{1}{2}Lm_1,
\end{equation*}
so that 
\begin{equation}\label{Hee}
H_{\ell}(L,M;n_1,\dots,n_k;q)
=q^{\frac{1}{4}mCm-\frac{1}{2}Lm_1}\prod_{j=1}^{k-1}\qbin{n_j+m_j}{n_j},
\end{equation}
where the $m_j$ follow from \eqref{eqp2c} and where the $n_j$ must satisfy
\begin{equation}\label{comp4}
2\sum_{j=1}^k j n_j=L+M+\ell-2k.
\end{equation}
Once again, to obtain the full generating function 
we need to sum over all admissible sequences $(n_1,\dots,n_k)$.
Since \eqref{Hee} has not explicit $n_k$ dependence and since
\eqref{comp4} may be rewritten as
\begin{equation*}
\frac{L+M+\ell}{2k}+(C^{-1}n)_1=1+n_1+n_2+\cdots+n_k
\end{equation*}
with $n=(n_1,\dots,n_{k-1})$,
this results in the right-hand side of \eqref{Q1}.

\subsection{Proof of Proposition~\ref{prop3}}\label{secprop3}

Let $\la=(\la_1,\la_2,\dots)$ be a partition, i.e., 
$\la_1\geq \la_2\geq \dots\geq 0$ with finitely many $\la_i$ unequal to zero.
The nonzero $\la_i$ are called the parts of $\la$.
The weight $\abs{\la}$ of $\la$ is the sum of its parts.
If $m_i(\la)=m_i$ is the multiplicity of the part $i$ in $\lambda$ we will
also write $\la=(1^{m_1},2^{m_2},\dots)$.
Given a partition $\lambda$ with largest part at most $k-1$ we define
\begin{equation*}
e_{\lambda}=e_{\lambda_1}+e_{\lambda_2}+\cdots=
\sum_{i=1}^{k-1} m_i(\lambda) e_i.
\end{equation*}

For $r$ an integer, $\sigma\in\{0,1\}$ and $\lambda$ a partition with 
largest part at most $k-1$ the following identity was proved 
in \cite[Cor. 4.1]{SW98}:
\begin{subequations}\label{SW}
\begin{multline}
\sum_{L=0}^{\infty}\frac{q^{L(L+r)/k}}{(q;q)_L(q;q)_{L+r}}
\sum_{\substack{n\in\Z^{k-1} \\
\frac{2L+r-\abs{\lambda}+k\sigma}{2k}-(C^{-1}n)_1\in\Z}}
q^{nC^{-1}(n-e_{\lambda})}\prod_{i=1}^{k-1}\qbin{n_i+m_i}{n_i} \\
=\frac{1}{(q;q)_{\infty}}
\sum_{\substack{n\in\Z^{k-1} \\
\frac{r-\abs{\lambda}+k\sigma}{2k}-(C^{-1}n)_1\in\Z}}
\frac{q^{nC^{-1}(n-e_{\lambda})}}{(q;q)_{n_1}\cdots(q;q)_{n_{k-1}}},
\end{multline}
where
\begin{equation}
m=C^{-1}\Bigl[(2L+r)e_{k-1}+e_{\lambda}-2n\Bigr]
\end{equation}
and 
\begin{equation}
r+\abs{\lambda}+\sigma k\equiv 0\pmod{2}.
\end{equation}
\end{subequations}
Note that $\sigma$ is fixed if $k$ is odd, but can be
either $0$ or $1$ when $k$ is even.

If we choose $\la$ and $\sigma$ as
\begin{equation*}
\la=(k-\ell,(k-1)^{n_0}) \qquad\text{and}\qquad \sigma\equiv n_0+1\pmod{2}
\end{equation*}
with $\ell\in\{1,\dots,k\}$ 
(when $\ell=1$ or $\ell=k$ the above should read $\la=((k-1)^{n_0+1})$ or
$\la=((k-1)^{n_0})$, respectively),
then \eqref{SW} becomes
\begin{subequations}\label{WS}
\begin{multline}\label{WS1}
\sum_{L=0}^{\infty}\frac{q^{L(L+r)/k}}{(q;q)_L(q;q)_{L+r}}
\sum_{\substack{n\in\Z^{k-1} \\
\frac{2L+r+n_0+\ell}{2k}-(C^{-1}n)_1\in\Z}}
q^{nC^{-1}(n-n_0e_{k-1}-e_{k-\ell})}\prod_{i=1}^{k-1}\qbin{n_i+m_i}{n_i} \\
=\frac{1}{(q;q)_{\infty}}
\sum_{\substack{n\in\Z^{k-1} \\
\frac{r+n_0+\ell}{2k}-(C^{-1}n)_1\in\Z}}
\frac{q^{nC^{-1}(n-n_0e_{k-1}-e_{k-\ell})}}{(q;q)_{n_1}\cdots(q;q)_{n_{k-1}}},
\end{multline}
where
\begin{equation}
m=C^{-1}\Bigl[(2L+r+n_0)e_{k-1}+e_{k-\ell}-2n\Bigr]
\end{equation}
and
\begin{equation}
n_0+r+\ell\equiv 0\pmod{2}.
\end{equation}
\end{subequations}

So far we have assumed that $\ell\in\{1,\dots,k\}$. 
If in \eqref{SW} we choose 
\begin{equation*}
\la=((k-1)^{n_0}) \qquad\text{and}\qquad \sigma\equiv n_0\pmod{2}
\end{equation*}
it easily follows that we obtain the $\ell=0$ instance of
\eqref{WS}.

The final step in our proof consists of changing the summation indices $n_i$ to
$n_{k-i}$ on both sides of \eqref{WS1} (on the left-hand side
we of course also change $m_i\to m_{k-i}$). Since
\begin{equation*}
(C^{-1}n)_1 + (C^{-1}n)_{k-1}\in\Z
\end{equation*}
this yields Proposition~\ref{prop3}.

\section{$\Bup=\Fup$ in the large $k$ limit}\label{sec4}

In the large $k$ limit our proof that $\Bup=\Sup=\Fup$ can be 
considerably simplified and made purely analytic, as 
will be demonstrated below.

In fact, since the equivalence between the spinon and fermionic
forms of the graded characters rests on Proposition~\ref{prop3}
which does not significantly simplify in the large $k$ limit,
we will only consider the $\Bup=\Fup$ correspondence here.

When $k$ tends to infinity the bosonic character $\chi_{\ell,r}^{\Bup}(q)$
of \eqref{carag} trivializes to
\begin{equation*}
V_r(q)-V_{r+2\ell+1}(q)
\end{equation*}
since only the singular vector of lowest conformal dimension yields a
non-vanishing contribution.
This should equate with the large $k$ limit of the 
fermionic character
$\chi_{\ell,r}^{\Fup}(q)$ of \eqref{fercar}.
\begin{theorem}
Let $\ell$ to be a positive integer and $r$ be any integer.
Then
\begin{multline}\label{BFinf}
\lim_{k\to\infty}\sum_{\substack{n_0,\dots,n_{k-1}=0 \\ 
\frac{n_0-r}{2k}+(C^{-1}n)_1\in\Z}}^{\infty}
\frac{q^{(r-n_0)(r+n_0+2\ell)/(4k)+\binom{n_0+1}{2}+n C^{-1}(n-n_0 
e_1-e_{\ell})}}
{(q;q)_{n_0}\cdots (q;q)_{n_{k-1}}} \\
=V_r(q)-V_{r+2\ell+1}(q).
\end{multline}
\end{theorem}
We note that for later convenience we have used that
$\chi_{\ell,r}(q)=\chi_{\ell,-r-2\ell}(q)$ on the 
left-hand side. Hence the restriction on the sum does not
exactly match that of \eqref{BF}.
We also remark that the summand on the left only makes sense for
$k\geq\ell$. Since we consider the large $k$ limit for fixed $\ell$
this is of course not a significant issue.

Before proving the above theorem we first introduce some more
notation relating to partitions (see Section~\ref{secprop3}).
The length $l(\la)$ of a partition $\la$
is the number of parts (non-zero $\la_i$).
If the weight of $\la$ is $n$, i.e., $\abs{\la}=n$ we write $\la\vdash n$ 
and say that $\la$ is a partition of $n$.
We identify a partition with its Ferrers graph,
defined by the set of points in $(i,j)\in \Z^2$ such that
$1\leq j\leq \lambda_i$.
The conjugate $\lambda'$ of $\lambda$ is the partition obtained by
reflecting the diagram of $\lambda$ in the main diagonal,
so that, in particular, $m_i(\la)=\la_i'-\la_{i+1}'$.

Given a partition we set
\begin{equation*}
n(\la)=\sum_{i\geq 1}(i-1)\la_i=\sum_{i\geq 1}\binom{\la_i'}{2}
\end{equation*}
and 
\begin{equation*}
b_{\la}(q)=\prod_{i\geq 1}(q;q)_{m_i(\la)}=
\prod_{i\geq 1}(q;q)_{\la'_i-\la'_{i+1}}.
\end{equation*}

In order to take the limit in \eqref{BFinf} we introduce two partitions
$\la$ and $\mu$ such that $l(\la)\leq \lfloor k/2\rfloor$ and
$l(\mu)\leq \lfloor (k-1)/2\rfloor$ as follows
\begin{gather*}
\la_i=n_i+n_{i+1}\cdots+n_{\lfloor k/2\rfloor}, \\
\mu_i=n_{k-i}+n_{k-i-1}\cdots+n_{1+\lfloor k/2\rfloor}.
\end{gather*}
For example, when $k=7$, $\la=(n_1+n_2+n_3,n_2+n_3,n_3)$, 
$\mu=(n_4+n_5+n_6,n_4+n_5,n_4)$ and conversely,
$n=(\la_1-\la_2,\la_2-\la_3,\la_3,\mu_3,\mu_2-\mu_3,\mu_1-\mu_2)$.

A simple calculation shows that for 
$\ell\in\{0,\dots,{\lfloor k/2\rfloor}\}$,
\begin{align*}
n C^{-1}&(n-n_0 e_1-e_{\ell}) \\
&=\sum_{i\geq 1} (\la_i^2+\mu_i^2)-n_0\la_1-(\la_1+\cdots+\la_{\ell})
-(\abs{\la}+\abs{\mu})(\abs{\la}+\abs{\mu}+n_0+\ell)/k \\
&=2n(\la')+2n(\mu')+\abs{\la}+\abs{\mu}-n_0\la_1
-(\la_1+\cdots+\la_{\ell}) \\[2mm]
& \qquad\qquad\qquad\qquad
-(\abs{\la}+\abs{\mu})(\abs{\la}+\abs{\mu}+n_0+\ell)/k.
\end{align*}
Also,
\begin{equation*}
(C^{-1}n)_1=\frac{1}{k}(\abs{\mu}-\abs{\la})+\la_1.
\end{equation*}
Using these two results the large $k$ limit can easily be taken
leading to
\begin{multline*}
\text{LHS}\eqref{BFinf}=
\sum_{\substack{n_0=0 \\ n_0\equiv r \ppmod{2}}} \;
\sum_{\substack{\la,\mu \\ \abs{\la}-\abs{\mu}=\frac{1}{2}(n_0-r)}} \\
\times
\frac{q^{\binom{n_0+1}{2}-n_0\la_1+2n(\la')+2n(\mu')+\abs{\la}+\abs{\mu}
-(\la_1+\cdots+\la_{\ell})}}
{(q;q)_{n_0}\, b_{\la'}(q)b_{\mu'}(q)}.
\end{multline*}
To further simplify this we invoke Hall's identity \cite{Hall}
\begin{equation}\label{hall}
\sum_{\lambda\vdash j} \frac{q^{2n(\la)}}
{b_{\la}(q)}=\frac{1}{(q;q)_j}
\end{equation}
(with $\la\to \mu'$) to find
\begin{multline*}
\text{LHS}\eqref{BFinf}=
\sum_{j=0}^{\infty}
\sum_{\substack{m=0 \\ m\equiv r \ppmod{2}}}^{\infty} \;
\sum_{\la\vdash j+\frac{1}{2}(m-r)}
\frac{q^{j+\binom{m+1}{2}-m\la_1+2n(\la')+\abs{\la}
-(\la_1+\cdots+\la_{\ell})}}
{(q;q)_j (q;q)_m b_{\la'}(q)},
\end{multline*}
where we have also replaced $n_0$ by $m$.
Key to showing that this is equal to the right-hand side of \eqref{BFinf}
is the identity
\begin{multline}\label{key}
\sum_{\substack{m=0 \\ m\equiv j \ppmod{2}}}^{\infty}\:
\sum_{\la\vdash\frac{1}{2}(j+m)}
\frac{q^{\binom{m+1}{2}-m\la_1
+2n(\la')+\abs{\la}-(\la_1+\dots+\la_{\ell})}}
{(q;q)_m b_{\la'}(q)} \\
=\frac{1}{(q;q)_{\infty}}
\sum_{i=0}^{\infty}
\frac{q^{\binom{j-2i}{2}}-q^{\binom{j-2i-2\ell-1}{2}}}{(q;q)_i},
\end{multline}
where $j$ and $\ell$ are integers such that $\ell\geq 0$.
Utilizing this with $j\to 2j-r$ leads to
\begin{align*}
\text{LHS}\eqref{BFinf}&=\frac{1}{(q;q)_{\infty}}
\sum_{i,j=0}^{\infty} \frac{q^{\binom{2j-2i-r}{2}+j}-
q^{\binom{2j-2i-2\ell-r-1}{2}+j}}{(q;q)_i(q;q)_j} \\
&=V_r(q)-V_{r+2\ell+1}(q) \\[2mm]
&=\text{RHS}\eqref{BFinf}
\end{align*}
as desired.

The rest of this section is devoted 
to proving \eqref{key}.
First we change the summation index $m$ to
$2m-j$, cancel a common factor $q^{\binom{j}{2}}$
and use
\begin{equation*}
\sum_{m=0}^{\infty} \sum_{\la\vdash m}f_{m,\la}\to
\sum_{\la}f_{\abs{\la},\la}.
\end{equation*}
We are then left with the $a=q^{-j}$ instance of
\begin{multline}\label{aid}
\sum_{\la}
\frac{a^{2\abs{\la}-\la_1}q^{\binom{2\abs{\la}+1}{2}-2\abs{\la}\la_1
+2n(\la')+\abs{\la}-(\la_1+\cdots+\la_{\ell})}}
{(aq;q)_{2\abs{\la}}\,b_{\la'}(q)} \\
=\frac{1}{(aq;q)_{\infty}}
\sum_{i=0}^{\infty}
\frac{a^{2i}q^{\binom{2i+1}{2}}
-a^{2i+2\ell+1}q^{\binom{2i+2\ell+2}{2}}}{(q;q)_i}.
\end{multline}
Our next step is to rename $\la_1$ as $k$ and to let $\mu$
be the partition $\mu=(\la_2,\la_3,\dots)$. Also denoting
$\abs{\mu}$ by $j$ this gives
\begin{equation*}
\text{LHS}\eqref{aid}=
\sum_{j,k=0}^{\infty}\sum_{\mu\vdash j}
\frac{a^{2j+k} q^{j(2j+2k+1)+k^2+k\delta_{\ell,0}
+2n(\mu')+\abs{\mu}-(\mu_1+\cdots+\mu_{\ell-1})}}
{(aq;q)_{2j+2k}(q;q)_{k-\mu_1}\,b_{\mu'}(q)} \
\end{equation*}
The sum over $\mu$ can now be performed by the following
identity from \cite{W04}:
\begin{equation}\label{w04}
\sum_{\lambda\vdash j}
\frac{q^{2n(\la')+\abs{\la}-\sum_{i=1}^m \lambda_i}
(q;q)_k}{(q;q)_{k-\la_1}b_{\la'}(q)} \
=\qbin{k+j-1}{j}-(1-q^k)\qbin{k+j-m-1}{j-m-1},
\end{equation}
for $m\in\{0,\dots,j\}$. We should remark that
the above assumes a slightly different definition of the $q$-binomial
coefficient than given in \eqref{qbinom}, namely
\begin{equation*}
\qbin{n+j}{j}=\begin{cases}
\displaystyle
\frac{(q^{n+1};q)_j}{(q;q)_j} & j\in\{0,1,2,\dots\} \\[3mm]
0 & \text{otherwise.}
\end{cases}
\end{equation*}
By \eqref{w04} the $\mu$-sum times $(q;q)_k$ gives
\begin{equation*}
\qbin{k+j-1}{j}-(1-q^k)\qbin{k+j-\ell}{j-\ell}
\end{equation*}
for $\ell>0$ and 
\begin{equation*}
\qbin{k+j-1}{j}-(1-q^k)\qbin{k+j-1}{j-1}
=q^{-k}\qbin{k+j-1}{j}-q^{-k}(1-q^k)\qbin{k+j}{j}
\end{equation*}
for $\ell=0$.
Combining the above two expressions we therefore find that
\begin{equation*}
\text{LHS}\eqref{aid}=
\sum_{j,k=0}^{\infty}
\frac{a^{2j+k} q^{j(2j+2k+1)+k^2}}
{(aq;q)_{2j+2k}(q;q)_k}
\biggl\{ \qbin{k+j-1}{j}-(1-q^k)\qbin{k+j-\ell}{j-\ell} \biggr\}.
\end{equation*}
This should be equated with the right-hand side of \eqref{aid}.
In fact, as we shall see, the following dissection takes place:
\begin{equation}\label{een}
\sum_{j,k=0}^{\infty}
\frac{a^{2j+k} q^{j(2j+2k+1)+k^2}}
{(aq;q)_{2j+2k}(q;q)_k} \qbin{k+j-1}{j}
=\frac{1}{(aq;q)_{\infty}}
\sum_{i=0}^{\infty}
\frac{a^{2i}q^{\binom{2i+1}{2}}}{(q;q)_i}
\end{equation}
and
\begin{equation*}
\sum_{j,k=0}^{\infty}
\frac{a^{2j+k} q^{j(2j+2k+1)+k^2}}
{(aq;q)_{2j+2k}(q;q)_{k-1}}
\qbin{k+j-\ell}{j-\ell}
=\frac{1}{(aq;q)_{\infty}}
\sum_{i=0}^{\infty}
\frac{a^{2i+2\ell+1}q^{\binom{2i+2\ell+2}{2}}}{(q;q)_i}.
\end{equation*}
Replacing $j\to j+\ell$, $k\to k+1$ and $a\to aq^{-2\ell-1}$
this last identity may also be stated as
\begin{equation}\label{twee}
\sum_{j,k=0}^{\infty}
\frac{a^{2j+k} q^{j(2j+2k+1)+k^2+k}}
{(aq;q)_{2j+2k+1}(q;q)_k} \qbin{k+j+1}{j}
=\frac{1}{(aq;q)_{\infty}}
\sum_{i=0}^{\infty}
\frac{a^{2i}q^{\binom{2i+1}{2}}}{(q;q)_i}
\end{equation}
independent of $\ell$.
Both \eqref{een} and \eqref{twee} are special cases of the
more general
\begin{equation}\label{drie}
\sum_{j,k=0}^{\infty}
\frac{a^{2j}b^k q^{j(j+1)+(j+k)^2}}
{(bq;q)_{2j+2k}(q;q)_k}\, \frac{(b^2q^k/a^2;q)_j}{(q;q)_j}
=\frac{1}{(bq;q)_{\infty}}
\sum_{i=0}^{\infty}
\frac{a^{2i}q^{\binom{2i+1}{2}}}{(q;q)_i}.
\end{equation}
To prove this we shift the summation index $k\to k-j$ and employ basic
hypergeometric notation to get
\begin{equation*}
\text{LHS}\eqref{drie}=
\sum_{k=0}^{\infty}
\frac{b^k q^{k^2}}
{(q;q)_k(bq;q)_{2k}}\:
{_2\phi_1}\biggl[\genfrac{}{}{0pt}{}{a^2q^{1-k}/b^2,q^{-k}}{0};
q,bq^{2k+1}\biggr].
\end{equation*}
By Heine's transformation \cite[Eq. (III.2)]{GR}
\begin{equation*}
{}_2\phi_1(a,b;c;q,z)=
\frac{(c/b,bz;q)_{\infty}}{(c,z;q)_{\infty}} \:
{_2\phi_1}(abz/c,b;bz;q,c/b)
\end{equation*}
this may be rewritten as
\begin{align*}
\text{LHS}\eqref{drie}&=
\sum_{k=0}^{\infty}
\frac{b^k q^{k^2}} {(q,bq;q)_k}\:
\lim_{\gamma\to 0}
{_2\phi_1}\biggl[\genfrac{}{}{0pt}{}{a^2q^2/b\gamma,q^{-k}}{bq^{k+1}};
q,\gamma q^k\biggr] \\
&=\sum_{k=0}^{\infty}\sum_{j=0}^k
\frac{a^{2j}b^{k-j} q^{j^2+j+k^2}} {(q;q)_j(q;q)_{k-j}(bq;q)_{j+k}}.
\end{align*}
Changing the order of the two sums and shifting $k\to k+j$ this
becomes
\begin{align*}
\text{LHS}\eqref{drie}&=
\sum_{j=0}^{\infty}
\frac{a^{2j} q^{2j^2+j}}{(q;q)_j(bq;q)_{2j}}\:
\lim_{\gamma,\delta\to \infty}
{_2\phi_1}\biggl[\genfrac{}{}{0pt}{}{\gamma,\delta}
{bq^{2j+1}};q,\frac{bq^{2j+1}}{\gamma\delta}\biggr] \\
&=\frac{1}{(bq;q)_{\infty}} \sum_{j=0}^{\infty}
\frac{a^{2j} q^{2j^2+j}} {(q;q)_j}
\end{align*}
in accordance with the right-hand side of \eqref{drie}.
To obtain the final expression on the right we have employed 
the $q$-Gauss sum \cite[Eq. (II.8)]{GR}
\begin{equation*}
{_2\phi_1}(a,b;c;q,c/ab)=\frac{(c/a,c/b;q)_{\infty}}{(c,c/ab;q)_{\infty}}.
\end{equation*}

\appendix

\section{The identity \eqref{bfB}}

Up to trivial manipulations, the difference between 
\eqref{string} and \eqref{cara} is due to a different
representation of the Verma character $\Vh_t(q)$.

Indeed, we note that \eqref{vermu} can be 
written in $q$-hypergeometric notation as
\begin{equation*}
\Vh_t(q)=\frac{q^{\max\{0,t\}}}{(q;q)_{\abs{t}}}\,
{_2\phi_1}(0,0;q^{\abs{t}+1};q,q)
\end{equation*}
By Heine's transformation \cite[Eq. (III.1)]{GR}
\begin{equation*}
{_2\phi_1}(a,b;c;q,z)=
\frac{(b,az;q)_{\infty}}{(c,z;q)_{\infty}} \; {_2\phi_1}(c/b,z;az;q,b)
\end{equation*}
this yields
\begin{equation*}
\Vh_t(q)=
\frac{q^{\max\{0,t\}}}{(q;q)^2_{\infty}} 
\sum_{i=0}^{\infty} (-1)^i q^{\binom{i+1}{2}+i\abs{t}}.
\end{equation*}
Since
\begin{equation}\label{nul}
\sum_{i=0}^{\infty} (-1)^i q^{\binom{i+1}{2}+it}=0
\end{equation}
for $t\in\Z$ this may be simplified to
\begin{equation}\label{Vbos}
\Vh_t(q)=\frac{1}{(q;q)^2_{\infty}} 
\sum_{i=0}^{\infty} (-1)^i q^{\binom{i+1}{2}-it}.
\end{equation}
It is this `bosonic' form for the Verma character
that is responsible for those minus signs in \eqref{string} that
are not due to the usual addition and subtraction of singular vectors.

Substituting \eqref{Vbos} in \eqref{cara} and using that
$2r=m-\ell$ we obtain
\begin{multline}
\hat{\chi}_{\ell,r}(q)
=\frac{1}{\eta^2(q)}
\sum_{j=-\infty}^{\infty}\sum_{i=0}^{\infty}(-1)^i 
q^{-(m+ik)^2/4k} \\
\times \Bigl\{
q^{\bigl(\ell+1+(i+2j)(k+2)\bigr)^2/4(k+2)}
-q^{\bigl(\ell+1-(i-2j)(k+2)\bigr)^2/4(k+2)}\Bigr\}.
\end{multline}
Replacing $j\to -j$ in the sum corresponding to the second term
in the summand and then splitting the sum over $j$ and using
\eqref{nul} in the form 
$\sum_{i\geq 0}\dots=-\sum_{i<0} \dots$
completes the proof of \eqref{bfB}.

We should remark that a similar kind of rewriting of the
graded verma character \eqref{gradeV} may be carried out to yield
the bosonic form
\begin{equation*}
V_t(q)=\frac{1}{(q;q)^3_{\infty}}
\sum_{j=-\infty}^{\infty}\sum_{i=0}^{\infty}
(-1)^i q^{\binom{2j-t}{2}+\binom{i+1}{2}+j(i+1)}.
\end{equation*}

\medskip

An alternative --- albeit somewhat indirect --- demonstration
of \eqref{bfB} is to show that \eqref{cara} follows from
\eqref{bfct} and \eqref{bchihat}. This is easily
achieved as follows.
By the Jacobi triple product identity \eqref{jaco} and the
expansion \eqref{keyexp} the 
denominator of \eqref{WK} may be put as 
\begin{equation*}
\frac{1}{\displaystyle\sum_{\sigma\in\{\pm 1\}}
\Theta^{(2)}_{\sigma}(x;q)}
=\frac{q^{-1/12}x^{1/2}}{\eta(q)}\sum_{n=-\infty}^{\infty}
x^{-n}\Vh_n(q),
\end{equation*}
with $\Vh_n(q)$ the character \eqref{vermu}.

Multiplying the above expression by the numerator of \eqref{WK} yields
\begin{multline*}
\hat{\chi}_{\ell}(x;q)
=\frac{q^{-1/12}}{\eta(q)}
\sum_{\sigma\in\{\pm 1\}}\sigma  \\
\times
\sum_{j,n=-\infty}^{\infty}
x^{-n-j(k+2)-\sigma(\ell+1)/2+1/2}
q^{(k+2)\Bigl(\sigma j+\tfrac{\ell+1}{2(k+2)}\Bigr)^2} \Vh_n(q).
\end{multline*}
Shifting $n\to n-(k+2)j-(\sigma-1)(\ell+1)/2$ 
and replacing $j$ by $\sigma j$ leads to
\begin{equation*}
\hat{\chi}_{\ell}(x;q)
=\frac{1}{\eta(q)}
\sum_{n=-\infty}^{\infty} x^{-n-\ell/2} q^{(n+\ell/2)^2/k}
\hat{\chi}_{\ell,n}^{\Bupp}(q)
\end{equation*}
with  $\hat{\chi}_{\ell,r}^{\Bupp}(q)$ given by \eqref{cara}.
Finally setting $n=kj+(m-\ell)/2$ with $1-k\leq m\leq k$ and using
the symmetry $\hat{\chi}_{\ell,r}(q)=\hat{\chi}_{\ell,r+jk}(q)$ yields
\begin{equation*}
\hat{\chi}_{\ell}(x;q)=\sum_{\substack{m=1-k \\ m-\ell \text{ even}}}^k
\hat{\chi}_{\ell,\tfrac{m-\ell}{2}}^{\Bupp}(q) K_m(x;q).
\end{equation*}

\end{document}